\documentclass[journal]{IEEEtran}
\usepackage[caption=false,font=normalsize,labelfont=rm,textfont=rm]{subfig}
\usepackage{amsmath,amsthm,amsfonts,amssymb,empheq}
\usepackage{array,soul,xcolor}
\usepackage{mathtools,cases,relsize}
\usepackage[T1]{fontenc}

\usepackage{textcomp,xpatch}
\usepackage{stfloats} 
\usepackage{multirow}
\usepackage{graphicx}
\usepackage{url,verbatim}
\usepackage{cite,balance,hyperref}
\usepackage{nicefrac, xfrac}
\usepackage{aligned-overset}
\usepackage{algpseudocode}
\usepackage{placeins}
\usepackage{float}
\usepackage{mathrsfs}
\usepackage{physics}
\usepackage{booktabs}
\usepackage{epstopdf}


\newtheoremstyle{colon}
  {3pt}
  {3pt}
  {\itshape}
  {}
  {\bfseries}
  {:}
  { }
  {\thmname{#1}\thmnumber{ #2}\thmnote{ (#3)}}
\theoremstyle{colon}
\makeatletter
\xpatchcmd{\proof}{\@addpunct{.}}{\@addpunct{:}}{}{}
\makeatother

\renewcommand{\Re}{\mathfrak{R}}

\newtheorem{proposition}{Proposition}

\newtheorem{lemma}{Lemma}
\usepackage{algorithm}

\makeatletter
\newcommand{\smallbullet}{} 
\DeclareRobustCommand\smallbullet{%
  \mathord{\mathpalette\smallbullet@{0.5}}%
}
\newcommand{\smallbullet@}[2]{%
  \vcenter{\hbox{\scalebox{#2}{$\m@th#1\bullet$}}}%
}
\makeatother



\AddToHook{env/subequations/before}{\ifhmode\unskip\fi}

\title{\fontsize{20}{26}\selectfont Weighted Sum-Rate Optimization of Near-Field STAR-RIS-Aided ISAC Systems Under Hardware Impairments}
\author{Huy T. Nguyen,~\IEEEmembership{Member,~IEEE},
        Nhu-Thinh Bui Doan,
        Minh-Tuan Pham,
        Thai-Hoc Vu,~\IEEEmembership{Member,~IEEE},\\
        Tien-Hoa Nguyen,~\IEEEmembership{Member,~IEEE},
        and Lajos Hanzo,~\IEEEmembership{Fellow,~IEEE}
\thanks{This research was funded by the Vietnam National Foundation for Science and Technology Development (NAFOSTED) under Grant number 102.04-2024.01. Corresponding author: Nguyen Tien Hoa (email: hoa.nguyentien@hust.edu.vn)}
\thanks{Huy T. Nguyen is with Smart and Autonomous Systems Research Group, Faculty of Information Technology, School of Technology, Van Lang University, Ho Chi Minh City, 70000, Vietnam. (e-mail: huy.nt@vlu.edu.vn)} 

\thanks{Nhu-Thinh Bui Doan, Minh-Tuan Pham, and Tien-Hoa Nguyen are with the School of Electrical and Electronic Engineering, Hanoi University of Science and Technology, Hanoi, Vietnam. (e-mails: \{thinh.bdn233660, tuan.pm233701\}@sis.hust.edu.vn} 

\thanks{Thai-Hoc Vu is with the Faculty of Electrical Engineering and Computer Science, VSB-Technical University of Ostrava, 17. Listopadu 2172/15, 708 00, Ostrava, Czechia (e-mail: thai.hoc.vu@vsb.cz)} 

\thanks{L. Hanzo is with the School of Electronics and Computer Science, University of Southampton, Southampton, SO17 1BJ, U.K. (email: lh@ecs.soton.ac.uk).}
}

\begin{document}
\bstctlcite{BSTcontrol}
\maketitle
\begin{abstract}
This paper investigates weighted sum-rate (WSR) optimization for a simultaneously transmitting and reflecting reconfigurable intelligent surface (STAR-RIS)-assisted multi-user multiple-input multiple-output (MIMO) integrated sensing and communication (ISAC) system operating under the energy-splitting (ES) protocol. A hybrid propagation model is considered, where the base station (BS)-STAR-RIS channel follows far-field propagation, whereas the STAR-RIS-user channels are modeled using near-field spherical wavefronts. Practical hardware impairments (HWIs), including STAR-RIS phase perturbations and signal-dependent additive distortion at the BS and user equipment (UEs), are incorporated into the system model. Under this framework, the WSR maximization problem is formulated subject to radar beampattern, quality-of-service, transmit-power, and STAR-RIS ES constraints. The resulting problem is highly non-convex due to the coupling among the active beamformers, passive STAR-RIS coefficients, and HWI-dependent signal-to-interference-plus-noise ratio expressions. An alternating optimization (AO)-based successive convex approximation framework is developed to iteratively solve the active and passive beamforming subproblems. Numerical results demonstrate stable convergence of the proposed algorithm and consistent WSR improvement over direct-link transmission. Besides, severe HWIs also lead to WSR saturation in the high transmit-power regime, whereas stringent sensing constraints reduce communication throughput. Increasing the STAR-RIS aperture and the number of BS antennas improves the achievable WSR, although the performance gain becomes limited under strong hardware distortion.
\end{abstract}

\section{Introduction}\label{sec1}

\IEEEPARstart{R}{econfigurable} intelligent surfaces (RISs) have attracted considerable attention as programmable wireless environments for improving coverage, interference management, and energy efficiency \cite{pan2020multicell, liu2021reconfigurable,ahmad2023reconf,ni2025reconfigurable}. Unlike conventional reflecting-only RISs, simultaneously transmitting and reflecting RIS (STAR-RIS) can simultaneously serve users on both sides of the surface, enabling full-space coverage \cite{Liu2021STARWireless, Xu2021STARRIS, Mu2022STAR, Liu2022Coupled}. This capability is particularly attractive in millimeter-wave and Terahertz (THz) systems, where severe blockage and penetration loss may weaken direct BS--UE links \cite{Liu2021STARWireless, Wang2022STARS}.

With large apertures and high carrier frequencies, practical STAR-RIS deployments may operate in the near-field region, where spherical-wave propagation must be considered instead of the conventional planar-wave model \cite{zhou2024near, li2023joint}. In this regime, the channel response depends jointly on direction and distance, enabling spatial beamfocusing and additional spatial degrees of freedom \cite{li2024near, zhou2024near, bjornson2020power, nearfield_properties}. However, such a focus is sensitive to practical hardware impairments (HWIs), including STAR-RIS phase perturbations and signal-dependent transceiver distortion. {Similarly, \cite{11119774} also reported on relevant programmable-surface sensitive to impairment in optical RIS systems, where nonlinear hardware effects were incorporated into joint transmission and surface optimization.}

STAR-RIS-assisted integrated sensing and communication (ISAC) further couples the BS beamformers and STAR-RIS coefficients because the transmitted waveform must simultaneously satisfy communication and sensing requirements \cite{isac_overview, wang2023nearfield, isac_joint_mimo, mimo_radar_probing, he2022energy, wang2024beamforming, singh2024pareto}. {Sensing-aware transmission and localization-oriented designs have been investigated in \cite{singh2024multi}, while reconfigurable holographic surfaces mounted on aerial platforms have recently been considered for full-duplex ISAC in \cite{11414122}.} Under the energy-splitting (ES) protocol, the passive transmission and reflection coefficients are additionally coupled through the shared incident-power constraint, making the joint design more challenging in the presence of HWIs.

Although STAR-RIS-assisted ISAC has been investigated in several recent works \cite{yigit2025hybrid, wang2024beamforming, wei2024starris_secure, chen2024star}, most studies assume ideal hardware. {A recent exception is \cite{11526942}, which considered energy-efficient STAR-RIS-aided ISAC beamforming under HWIs using a foundation-model-based approach.} Existing RIS-assisted mmWave studies have also outlined hybrid beamforming and reflection-coefficient optimization \cite{singh2023energy, singh2024joint}. In practice, however, large passive surfaces and high-frequency RF chains are affected by phase perturbations and signal-dependent additive distortion \cite{papazafeiropoulos2023max, pang2025coverage}, whose impact becomes more pronounced at high transmit powers and large apertures \cite{li2023achievable, luo2024sum, mateos2025model}. Hence, the combined effects of hybrid far-/near-field propagation, ES-based STAR-RIS operation, ISAC sensing constraints, and practical HWIs remain uncovered.

\begin{table}[!t]
\centering
\caption{Comparison of the proposed work with representative studies.}
\label{tab:contrasting_table}

\renewcommand{\arraystretch}{1.12}
\setlength{\tabcolsep}{1pt}

\begin{tabular}{p{2.7cm}cccccccc}
\toprule
\textbf{Feature}
& \textbf{Our work}
& \cite{gupta2024affine}
& \cite{singh2023energy}
& \cite{singh2024joint}
& \cite{singh2024multi}
& \cite{singh2024pareto}
& \cite{mu2024reconfigurable}
& \cite{tang2023near} \\
\hline

RIS
& \checkmark & -- & \checkmark & \checkmark & -- & -- & \checkmark & \checkmark \\
\hline

STAR-RIS with ES protocol
& \checkmark & -- & -- & -- & -- & -- & P & -- \\
\hline

Hybrid far-/near-field propagation
& \checkmark & -- & -- & -- & -- & -- & \checkmark & \checkmark \\
\hline

ISAC / sensing-aware transmission
& \checkmark & P & -- & -- & P & P & P & -- \\
\hline

Joint active/passive beamforming
& \checkmark & -- & P & \checkmark & -- & -- & P & -- \\
\hline

Hardware impairments
& \checkmark & -- & -- & -- & -- & -- & -- & -- \\
\hline\hline
\multicolumn{9}{p{\columnwidth}}{\textit{Note:} ``P'' denotes partial consideration of the corresponding feature. ``--'' indicates that the feature is not explicitly considered.}
\\
\hline
\bottomrule
\end{tabular}
\end{table}

Against this background, this paper investigates sensing-constrained WSR maximization for a STAR-RIS-assisted downlink multi-user ISAC system under practical HWIs. The BS--STAR-RIS link follows far-field propagation, whereas the STAR-RIS--UE links are modeled using near-field spherical wavefronts. STAR-RIS phase perturbations and signal-dependent BS/UE distortion are incorporated into the system model, and the active BS beamformers and passive STAR-RIS transmission/reflection coefficients are jointly optimized subject to transmit-power, QoS, radar beampattern, and ES constraints. To address the non-convex problem, we employ an AO-based SCA framework with problem-specific convex surrogates for the active and passive subproblems.

Table~\ref{tab:contrasting_table} compares the proposed work with RIS-assisted ISAC and beamforming studies \cite{gupta2024affine, singh2023energy, singh2024joint, singh2024multi, singh2024pareto}. The considered framework jointly incorporates ES-based STAR-RIS operation, hybrid far-/near-field propagation, sensing constraints, and practical HWIs within a unified multi-user ISAC design. 
The main contributions are summarized as follows:
\begin{itemize}
\item We develop a STAR-RIS-assisted multi-user MIMO ISAC model that jointly captures ES operation, hybrid far-/near-field propagation, STAR-RIS phase perturbations, and signal-dependent BS/UE distortion.

\item We formulate a sensing-constrained WSR maximization problem that jointly optimizes the active BS beamformers and passive STAR-RIS transmission/reflection coefficients under transmit-power, QoS, radar beampattern, and ES constraints.
{
\item Based on AO and SCA techniques, we develop a problem-specific optimization framework that constructs HWI-aware concave rate surrogates, introduces auxiliary variables for BS distortion, derives an affine lower bound for the sensing constraint, and employs a penalty-based treatment of the non-convex equality ES constraint.}

\item Numerical results under different HWI levels, sensing thresholds,
STAR-RIS sizes, BS antenna configurations, and user loads demonstrate
stable convergence and the WSR advantage of the proposed design. The
results further show that severe HWIs can lead to WSR saturation at
high transmit powers, while more stringent sensing requirements reduce
communication throughput and larger STAR-RIS and BS arrays provide
performance gains subject to hardware-quality limitations.
\end{itemize}

\textit{Notations:}
Scalars, vectors, matrices, and sets are denoted by italic letters, boldface lower-case letters, boldface upper-case letters, and calligraphic letters, respectively. For a matrix $\mathbf{A}$, $\mathbf{A}^{T}$, $\mathbf{A}^{*}$, $\mathbf{A}^{H}$, and $\mathrm{Tr}(\mathbf{A})$ denote the transpose, conjugate, Hermitian transpose, and trace, respectively. $\mathbf{I}$ is the identity matrix. The operators $\mathrm{diag}(\mathbf{a})$ and $\mathrm{diag}(\mathbf{A})$ construct a diagonal matrix from $\mathbf{a}$ and extract the diagonal entries of $\mathbf{A}$, respectively. The Euclidean norm and absolute value are denoted by $\|\cdot\|$ and $|\cdot|$, respectively. $\mathbb{E}\{\cdot\}$, $\Re\{\cdot\}$, and $\otimes$ denote the expectation, real-part, and Kronecker-product operators, respectively. $\mathbb{C}^{M\times N}$ denotes the space of complex-valued $M\times N$ matrices, and $\triangleq$ means ``defined as''. A circularly symmetric complex Gaussian random vector with covariance $\mathbf{\Sigma}$ is denoted by $\mathcal{CN}(\mathbf{0},\mathbf{\Sigma})$.
\section{System Model}\label{sec2}

\begin{figure}[!t]
 \includegraphics[width=1.0\linewidth]{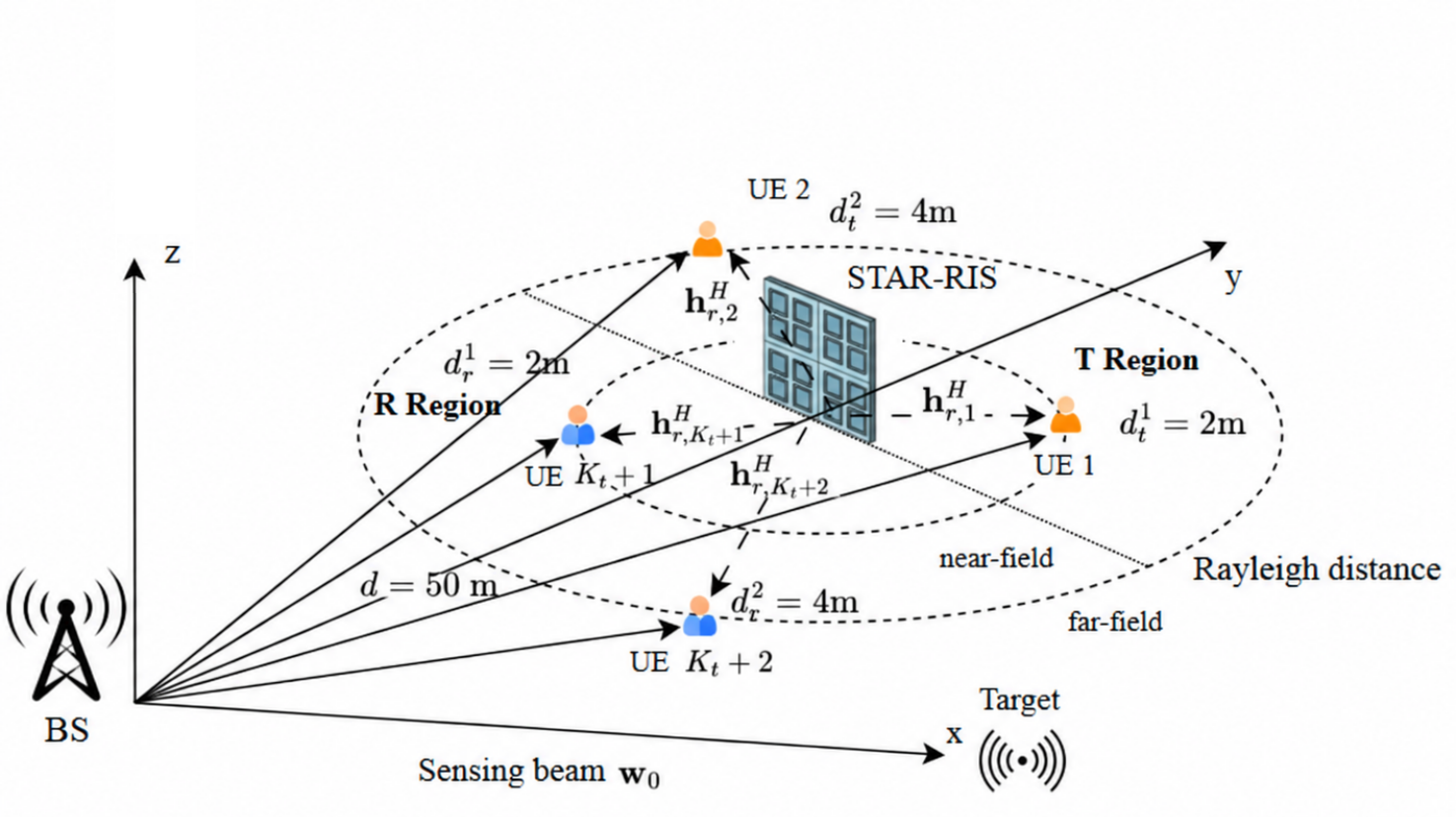}                            
        \caption{Illustration of the distinction between planar waves (far-field) and spherical waves (near-field) in the STAR-RIS system.}
            \label{fig:placeholder}
\end{figure}

As shown in Fig.~\ref{fig:placeholder}, we consider a STAR-RIS-assisted downlink multi-user MIMO system, where the system considered adopts a hybrid-field channel architecture determined by the Rayleigh distance criterion. Due to the relatively long BS-STAR-RIS separation, the BS-STAR-RIS channel is modeled using far-field planar-wave propagation, whereas the STAR-RIS-UE channels follow a near-field spherical-wave model. Unlike conventional far-field propagation, the near-field array response depends jointly on the angular direction and propagation distance. Consequently, the resultant channel structure can provide additional spatial DoFs for user separation and spatial beamfocusing. Furthermore, direct BS-UE links are modeled according to their corresponding propagation regions determined by the Rayleigh distance. In addition to communication links, a radar sensing channel is incorporated to impose beampattern constraints toward the sensing targets. Practical HWIs, including STAR-RIS phase perturbations and signal-dependent transceiver distortion at the BS and UEs, are incorporated directly into the system model.
\begin{table*}[!t]
\centering
\caption{List of Main Notations}
\label{tab:notations}
\renewcommand{\arraystretch}{1.1}
\footnotesize
{
\begin{tabular*}{\textwidth}{@{\extracolsep{\fill}}llllll@{}}
\toprule
\textbf{Symbol} & \textbf{Description} &
\textbf{Symbol} & \textbf{Description} &
\textbf{Symbol} & \textbf{Description} \\
\hline\hline

$M_b$ & Number of BS antennas. &
$\mathbf G$ & BS--STAR-RIS channel matrix. &
$\boldsymbol{\eta}_{\mathrm{BS}}$ & BS transmitter distortion. \\

$N$ & Number of STAR-RIS elements. &
$\mathbf h_k$ & STAR-RIS--UE $k$ channel. &
$\eta_{\mathrm{UE},k}$ & UE receiver distortion. \\

$\mathcal K$ & Set of all UEs. &
$\mathbf g_k$ & Direct BS--UE $k$ channel. &
$\mathbf\Upsilon_{\mathrm{BS}}$ & BS distortion covariance matrix. \\

$\mathcal K_t,\mathcal K_r$ & T-mode and R-mode UE sets. &
$\widetilde{\mathbf h}_k^{\mathrm{HWI}}$ &
HWI-aware effective UE channel. &
$\upsilon_{\mathrm{UE},k}$ & UE distortion variance. \\

$\mathbf x$ & Composite transmit signal. &
$p(\theta)$ & Beampattern gain at $\theta$. &
$f_{\mathrm{ES}}$ & Equality-ES penalty function. \\

$\mathbf v_k,\mathbf v_0$ &
Communication/sensing beamformers. &
$\Gamma_{\mathrm{tar}}$ & Minimum sensing gain. &
$\mu$ & ES penalty coefficient. \\

$\boldsymbol{\Phi}_i$ &
Nominal STAR-RIS matrix ($i=t,r$). &
$r_{\mathrm{qos}}$ & Minimum UE rate. &
$e_{\mathrm{ES}}^{(i)}$ & Equality-ES feasibility residual. \\

$\widetilde{\boldsymbol{\Phi}}_i$ &
Phase-perturbed STAR-RIS matrix. &
$\eta_k$ & UE priority weight. &
$\varepsilon_{\mathrm{ES}}$ & ES feasibility tolerance. \\

$\rho_n^i$ & Element-$n$ power-splitting coefficient. &
$\gamma_k^{\mathrm{HWI}},\gamma_k$ & HWI/ideal SINRs. &
$f_c$ & Carrier frequency. \\

$\theta_n^i$ & Element-$n$ nominal phase shift. &
$R_k^{\mathrm{HWI}},R_k$ & HWI/ideal achievable rates. &
$\lambda_c,k_c$ & Wavelength and wave number. \\

$\delta_n^i$ & Element-$n$ phase perturbation. &
$\kappa_{\mathrm{BS}},\kappa_{\mathrm{UE}}$ &
BS/UE HWI coefficients. &
$d_B,d_R$ & BS antenna/RIS element spacings. \\

$\mathbf q_t,\mathbf q_r$ & STAR-RIS T/R coefficient vectors. &
$\sigma_k^2$ & UE thermal-noise variance. &
$Z$ & Rayleigh distance. \\
\bottomrule
\end{tabular*}
}
\end{table*}

\subsection{STAR-RIS-Assisted Network Model}   
In the considered system as illustrated in Fig.~\ref{fig:placeholder}, the BS employs a uniform linear array (ULA) having $M_b$ antennas to simultaneously serve multiple UEs. In blockage-sensitive mmWave and THz environments, direct BS-UE links may experience severe attenuation due to limited scattering and penetration loss. To improve wireless coverage, a STAR-RIS operating under the ES protocol is deployed near the UEs. The STAR-RIS is modeled as a uniform planar array (UPA) composed of passive elements $N=N_yN_z$, and its set of element indexes is denoted by $\mathcal{N}\triangleq\{1,2,\ldots,N\}$.

UEs located on the transmission side of the STAR-RIS are referred to as T-UEs, whose index set is denoted by $\mathcal{K}_t \triangleq \{1,2,\ldots,K_t\}$. The remaining UEs located on the reflection side are referred to as R-UEs, with index set $\mathcal{K}_r \triangleq \{K_t+1,\ldots,K\}$. Accordingly, the overall UE index set is given by $\mathcal{K}=\mathcal{K}_t\cup\mathcal{K}_r$, where $\mathcal{K}_t\cap\mathcal{K}_r=\emptyset$ and $K_t+K_r=K$.

Under the ES protocol, each STAR-RIS element simultaneously supports transmission and reflection through coupled amplitude coefficients. The transmission and reflection responses are represented by diagonal matrices
\[
\boldsymbol{\Phi}_i=
\mathrm{diag} \Big(
\sqrt{\rho_1^i}e^{j\theta_1^i},
\sqrt{\rho_2^i}e^{j\theta_2^i},
\ldots,
\sqrt{\rho_N^i}e^{j\theta_N^i}
\Big), \quad i\in\{t,r\},
\]
where $\rho_n^t$ and $\rho_n^r$ denote the transmission and reflection amplitude coefficients of the $n$-th STAR-RIS element, respectively, while $\theta_n^t$ and $\theta_n^r$ represent the corresponding phase shifts.
In this work, the STAR-RIS operates under the ES protocol, where each passive element simultaneously supports transmission and reflection. Accordingly, the transmission and reflection coefficients of the $n$-th STAR-RIS element satisfy
\begin{align}
    \rho_n^t,\rho_n^r\in[0,1],  \rho_n^t+\rho_n^r=1,  \theta_n^t,\theta_n^r\in[0,2\pi), \, \forall n\in\mathcal N.
\end{align}
The ES protocol introduces coupled passive beamforming variables, since the transmission and reflection responses share the same incident power budget at each STAR-RIS element.

\subsection{Communication and Channel Model}
Typically, the BS's transmitted signal can be expressed as
\begin{equation} \label{dofs}
\mathbf{x}
=
\sum_{k\in\mathcal{K}}\mathbf{v}_k s_k
+
\mathbf{v}_0 s_0,
\end{equation}
with $\mathbf{x}\in\mathbb{C}^{M_b\times1}$. For each UE $k\in\mathcal{K}$, $s_k\in\mathbb{C}$ and $\mathbf{v}_k\in\mathbb{C}^{M_b\times1}$ denote the information symbol and transmit beamforming vector associated with UE $k$, respectively. In addition, $s_0\in\mathbb{C}$ and $\mathbf{v}_0\in\mathbb{C}^{M_b\times1}$ denote the radar probing signal and its corresponding beamforming vector.

\subsubsection{Far-Field BS-STAR-RIS Channel}
The BS--STAR-RIS channel is modeled by an $L$-path geometric one \cite{li2023joint} as
\begin{equation}
\mathbf{G}
=
\sqrt{\frac{\beta M_b N}{L}}
\sum_{l=1}^{L}
\mathbf{e}_{\mathrm{STAR}}(\varphi_l,\vartheta_l)
\mathbf{e}_{\mathrm{BS}}^{H}(\gamma_l),
\end{equation}
where $L$ and $\beta$ denote the number of propagation paths and the path-loss coefficient, respectively.

The STAR-RIS is equipped with a $N_y\times N_z$ UPA. Its array response vector is given by
\begin{align}
    \mathbf{e}_{\mathrm{STAR}}(\varphi_l,\vartheta_l)
=
\mathbf{e}_{y}(\varphi_l,\vartheta_l)
\otimes
\mathbf{e}_{z}(\vartheta_l),
\end{align}
where
\(
\mathbf{e}_{y}(\varphi_l,\vartheta_l)
=
[
1,
e^{j\Phi_{l,y}},
\ldots,
e^{j(N_y-1)\Phi_{l,y}}
]^T
\)
and
\(
\mathbf{e}_{z}(\vartheta_l)
=
[
1,
e^{j\Phi_{l,z}},
\ldots,
e^{j(N_z-1)\Phi_{l,z}}
]^T.
\)
The phase shifts are defined as
\(
\Phi_{l,y}=k_cd_R\sin\varphi_l\sin\vartheta_l\)
and
\(
\Phi_{l,z}=k_cd_R\cos\vartheta_l,
\)
where $k_c=2\pi/\lambda_c$ is the wave number, $\lambda_c$ is the carrier wavelength, and $d_R$ denotes the STAR-RIS element spacing. The variables $\varphi_l$ and $\vartheta_l$ represent the azimuth and elevation angles of arrival, respectively.

The BS employs a ULA with array response vector \cite{balanis2016antenna}
\[
\mathbf{e}_{\mathrm{BS}}(\gamma_l)
=
[
1,
e^{jk_c d_B\cos\gamma_l},
\ldots,
e^{jk_c(M_b-1)d_B\cos\gamma_l}
]^T,
\]
where $d_B=\lambda_c/2$ denotes the spacing of the BS antenna and $\gamma_l$ is the angle of departure associated with the BS.

\subsubsection{Near-Field STAR-RIS-UE LoS Channel}
The STAR-RIS is deployed on the $YZ$-plane, whereas the UEs are distributed on both sides of the surface in the $XY$-plane. The Cartesian coordinate of the $n$-th STAR-RIS element is given by \cite{li2023joint}
\begin{equation}
\mathbf{p}_n
=
\left[
0,\,
i_y(n)d_R+y_f,\,
i_z(n)d_R+z_f
\right]^T,
\end{equation}
where
$i_y(n)=\mathrm{mod}(n-1,N_y)$, and$\quad
i_z(n)=\left\lfloor(n-1)/N_y\right\rfloor$ denote the column and row indices of the $n$-th STAR-RIS element, respectively. The coordinate of UE $k$ is expressed as
\begin{align}
    \mathbf{u}_k
=
[x_{u,k},\,y_{u,k},\,0]^T.
\end{align}

Since near-field propagation is modeled using spherical wavefronts associated with free-space attenuation \cite{bjornson2020power}, the near-field LoS channel between UE $k$ and the $n$-th STAR-RIS element can be modeled as
\begin{align}
    [\mathbf{h}_k]_n
=
a_n^k
e^{-j\frac{2\pi r_n^k}{\lambda_c}},
\end{align}
where $a_n^k=\lambda_c/(4\pi r_n^k)$ is the free-space amplitude attenuation factor and $r_n^k=\|\mathbf{u}_k-\mathbf{p}_n\|_2$ is the propagation distance.

\subsubsection{Far-Field Sensing Beampattern}
The transmit beamforming design of the transmission jointly supports the communication and radar detection through directional beam pattern control. Accordingly, the transmit beampattern gain in angular direction $\theta$ can be formulated as in \cite{he2022energy} as
\begin{equation}
\begin{aligned}
p(\theta)
&=
\mathbb{E}\!\left\{
\left|
\mathbf{a}^{H}(\theta)\mathbf{x}
\right|^{2}
\right\}=
\mathbf{a}^{H}(\theta)
\left(
\sum_{k=1}^{K}\mathbf{v}_{k}\mathbf{v}_{k}^{H}
+
\mathbf{v}_{0}\mathbf{v}_{0}^{H}
\right)
\mathbf{a}(\theta),
\end{aligned}
\end{equation}
where $\mathbf{a}(\theta)=\frac{1}{\sqrt{M_b}}
[
1,
e^{j2\pi\frac{d_B}{\lambda_c}\sin\theta},
\ldots,
e^{j2\pi\frac{d_B}{\lambda_c}(M_b-1)\sin\theta}
]^T\in\mathbb{C}^{M_b\times1}$ is the steering vector associated with direction $\theta$.

\subsubsection{Direct BS-UE Channel}
The direct BS-UE channel depends on the propagation distance and angular direction between the BS and UE $k$. Let
$Z=(2D^{2})/{\lambda_c}$
denote the Rayleigh distance of the BS array, where $D=M_bd_B$ is the BS array aperture. Since the BS-UE separation considered is larger than the Rayleigh distance, the direct BS-UE link is modeled using the far-field approximation, i.e., $r_k>Z$.
Thus, the direct BS-UE channel $\mathbf{g}_k\in\mathbb{C}^{M_b\times1}$ is modeled using the far-field planar-wave approximation \cite{balanis2016antenna, zhang2023mixed}
\begin{equation}
\mathbf{g}_k^{H}
=
\sqrt{M_b}\alpha_k
\mathbf{a}_k^{H}(\theta_k),
\end{equation}
where 
\(
\alpha_k
=
{\sqrt{\beta}}
e^{-j{2\pi r_k}/{\lambda_c}}/{r_k}
\)
is the complex channel coefficient, with $\beta$ denoting the reference path-loss coefficient at a distance of 1 m, and the steering vector is given by
\(
\mathbf{a}_k(\theta_k)
=
\frac{1}{\sqrt{M_b}}
[
1,r
e^{j\pi\theta_k},
\ldots,
e^{j\pi(M_b-1)\theta_k}
]^T.
\)

\subsection{Hardware Impairment Model}
\label{sec_hwi}

{


This subsection incorporates practical HWIs into the considered STAR-RIS-assisted ISAC system, including element-wise phase perturbations at the STAR-RIS and signal-dependent additive distortion at the BS transmitter and UE receivers. These established models are adopted from \cite{papazafeiropoulos2023max}, without introducing a new hardware-error model. Although the underlying impairments are independent of the propagation regime, their effects in the considered near-field system depend on the spherical-wave STAR-RIS--UE channels. Specifically, element-dependent propagation distances influence both the phase coherence required for near-field focusing and the distortion observed at each UE, thereby coupling the adopted HWI models with the near-field beamfocusing process.
}

\subsubsection{STAR-RIS Phase Perturbation Model}
\label{subsec_ris_hwi}

Due to finite-resolution phase tuning and circuit imperfections,
the STAR-RIS coefficients are affected by random phase
perturbations. For each operation mode $i\in\{t,r\}$, the
practical STAR-RIS coefficient matrix is modeled as
\cite{papazafeiropoulos2023max}
\begin{equation}
\widetilde{\boldsymbol{\Phi}}_i
=
\mathrm{diag}
\!\left\{
\sqrt{\rho_1^i}e^{j(\theta_1^i+\delta_1^i)},
\dots,
\sqrt{\rho_N^i}e^{j(\theta_N^i+\delta_N^i)}
\right\}.
\label{eq:phi_hwi}
\end{equation}
{
%
%
The perturbation $\delta_n^i$ represents the residual phase error of the $n$-th STAR-RIS element caused by finite-resolution phase tuning and circuit imperfections \cite{papazafeiropoulos2023max}. Although its statistical model is independent of the propagation regime, its impact is particularly significant in near-field transmission, where the desired phase profile must compensate for the element-dependent distance $r_n^k$. Consequently, the actual phase $\theta_n^i+\delta_n^i$ introduces a mismatch in this distance-dependent compensation, thereby degrading coherent beamfocusing at UE $k$.
}

\subsubsection{Transceiver Distortion Model}
\label{subsec_trx_hwi}


{
Following \cite{papazafeiropoulos2023max}, we model the residual BS and UE hardware distortion as signal-dependent additive distortion that captures uncompensated RF imperfections, including quantization effects, I/Q imbalance, amplifier nonlinearities, and circuit imperfections. The aggregate distortion follows a zero-mean complex Gaussian distribution, with covariance proportional to the corresponding signal power and thus becoming more pronounced at high transmit power. Although its generation mechanism is independent of the propagation regime, the received distortion power depends on the effective channel, thereby coupling it with the spherical-wave STAR-RIS--UE channel and the resulting near-field beamfocusing gain.}

\subsubsection{BS transmitter distortion}

From the transmit signal $\mathbf{x}$ in \eqref{dofs},
the practical transmitted signal is modeled as
\cite{papazafeiropoulos2023max}
\begin{equation}
\label{eq:x_tilde}
\widetilde{\mathbf{x}}
=
\mathbf{x}
+
\boldsymbol{\eta}_{\mathrm{BS}},
\end{equation}
where
\(
\boldsymbol{\eta}_{\mathrm{BS}}
\sim
\mathcal{CN}
\!\left(
\mathbf{0},
\mathbf{\Upsilon}_{\mathrm{BS}}
\right)
\)
denotes the BS distortion noise. The covariance matrix is given
by
\(
\mathbf{\Upsilon}_{\mathrm{BS}}
=
\kappa_{\mathrm{BS}}^2
\operatorname{diag}(\mathbf{W}),
\)
where $\kappa_{\mathrm{BS}}^2\geq0$ denotes the BS impairment
coefficient, while
\[
\mathbf{W}
\triangleq
\mathbb{E}\{\mathbf{x}\mathbf{x}^H\}
=
\sum_{l\in\mathcal{K}}
\mathbf{v}_l\mathbf{v}_l^H
+
\mathbf{v}_0\mathbf{v}_0^H.
\]
This model captures the aggregate residual distortion induced
by practical radio-frequency HWIs.

{
To capture how the BS distortion propagates
through the STAR-RIS channel, we define the
channel of UE $k$ as
\begin{equation}
(\widetilde{\mathbf{h}}_k^{\mathrm{HWI}})^H
\triangleq
\mathbf{g}_k^H
+
\mathbf{h}_k^H
\widetilde{\boldsymbol{\Phi}}_i
\mathbf{G},
\qquad k\in\mathcal{K}_i.
\label{eq:h_eff_hwi}
\end{equation}
The BS distortion component observed at UE $k$ is therefore
$
d_{\mathrm{BS},k}
\triangleq
(\widetilde{\mathbf{h}}_k^{\mathrm{HWI}})^H
\boldsymbol{\eta}_{\mathrm{BS}}$. 
Given the effective channel, the linear transformation
of the Gaussian distortion vector writes
\begin{equation}
d_{\mathrm{BS},k}
\sim
\mathcal{CN}
\!\left(
0,
\sigma_{\mathrm{BS},k}^{2}
\right),
\label{eq:bs_dist_distribution}
\end{equation}
where $\sigma_{\mathrm{BS},k}^{2}
=
(\widetilde{\mathbf{h}}_k^{\mathrm{HWI}})^H
\mathbf{\Upsilon}_{\mathrm{BS}}
\widetilde{\mathbf{h}}_k^{\mathrm{HWI}}$ can be further derived as
\begin{align}
\sigma_{\mathrm{BS},k}^{2}
&=\!
\kappa_{\mathrm{BS}}^{2}
\sum_{m=1}^{M_b}
|
[\widetilde{\mathbf{h}}_k^{\mathrm{HWI}}]_m
|^{2} 
\Big[
\sum_{l\in\mathcal{K}}
|[\mathbf{v}_l]_m|^{2}
+
|[\mathbf{v}_0]_m|^{2}
\Big].\!\hspace{-3pt}
\label{eq:bs_dist_variance}
\end{align}
Hence, the BS distortion observed at UE $k$ jointly depends on the per-antenna transmit powers and near-field effective channel gains, thereby coupling active beamforming, STAR-RIS configuration, and BS hardware distortion.
}

\subsubsection{UE receiver distortion}

The receiver distortion at UE $k$ is modeled as
\cite{papazafeiropoulos2023max}
\begin{equation}
\label{eq:eta_ue}
\eta_{\mathrm{UE},k}
\sim
\mathcal{CN}
\!\left(
0,
\upsilon_{\mathrm{UE},k}
\right),
\end{equation}
{where the UE distortion variance $\upsilon_{\mathrm{UE},k}$ can be equivalently
expanded using the definition of $\mathbf{W}$ as
\begin{align}
\upsilon_{\mathrm{UE},k}
&=
\kappa_{\mathrm{UE}}^2
(\widetilde{\mathbf{h}}_k^{\mathrm{HWI}})^H
\mathbf{W}
\widetilde{\mathbf{h}}_k^{\mathrm{HWI}} \nonumber\\
&=
\kappa_{\mathrm{UE}}^2
\big(
\sum_{l\in\mathcal{K}}
\big|
(\widetilde{\mathbf{h}}_k^{\mathrm{HWI}})^H
\mathbf{v}_l
\big|^2
+
\big|
(\widetilde{\mathbf{h}}_k^{\mathrm{HWI}})^H
\mathbf{v}_0
\big|^2
\big),
\label{eq:ue_dist_expanded}
\end{align}
with $\kappa_{\mathrm{UE}}^2\geq0$ being the UE impairment
coefficient. Herein, $\upsilon_{\mathrm{UE},k}$ is proportional to the
aggregate received signal power associated with all
communication streams and the sensing stream. Unlike the BS
distortion in \eqref{eq:bs_dist_variance}, which depends on
the per-antenna transmit powers through
$\operatorname{diag}(\mathbf{W})$, the UE distortion depends
on the total received signal power after propagation through
the effective STAR-RIS-assisted channel.
}

The distortion terms $\boldsymbol{\eta}_{\mathrm{BS}}$ and
$\eta_{\mathrm{UE},k}$ are assumed to be independent of the
transmitted signals and thermal noise. Since both distortion
terms scale with the transmit covariance matrix $\mathbf{W}$,
the aggregate impairment power increases with the transmit
power. Consequently, the achievable SINR does not scale
linearly with $P_{\max}$, leading to the WSR saturation
observed under severe HWIs. {
Importantly, the transceiver distortion laws themselves remain
unchanged between far- and near-field propagation. The
near-field dependence enters through
$\widetilde{\mathbf{h}}_k^{\mathrm{HWI}}$ in
\eqref{eq:bs_dist_variance} and
\eqref{eq:ue_dist_expanded}. Thus, the adopted HWI model
naturally captures the interaction between signal-dependent
hardware distortion and distance-dependent near-field
beamfocusing.
}

\subsubsection{HWI-Aware Received Signal and SINR}
Under HWI models, the practical
signal received at UE $k$ is given by
\begin{equation}\label{eq:y_hwi}
\begin{split}
y_k^{\mathrm{HWI}}
&=
(\widetilde{\mathbf{h}}_k^{\mathrm{HWI}})^H
\widetilde{\mathbf{x}}
+
\eta_{\mathrm{UE},k}
+
n_k
\\
&=
(\widetilde{\mathbf{h}}_k^{\mathrm{HWI}})^H
\big(
\mathbf{x}
+
\boldsymbol{\eta}_{\mathrm{BS}}
\big)
+
\eta_{\mathrm{UE},k}
+
n_k,
\end{split}
\end{equation}
where $n_k\sim\mathcal{CN}(0,\sigma_k^2)$ denotes the
thermal noise. Accordingly, the effective SINR at UE $k$ is
expressed as
\begin{equation}\label{eq:sinr_hwi}
\gamma_k^{\mathrm{HWI}}
=
\big|
(\widetilde{\mathbf{h}}_k^{\mathrm{HWI}})^H
\mathbf{v}_k
\big|^2/
I_k^{\mathrm{HWI}}
,
\end{equation}
where
\begin{equation}\label{eq:Ik_hwi}
\begin{aligned}
I_k^{\mathrm{HWI}}
&=
\sum_{l\in\mathcal{K}\setminus\{k\}}
\big|
(\widetilde{\mathbf{h}}_k^{\mathrm{HWI}})^H
\mathbf{v}_l
\big|^2
+
\big|
(\widetilde{\mathbf{h}}_k^{\mathrm{HWI}})^H
\mathbf{v}_0
\big|^2
\\
&\quad+
(\widetilde{\mathbf{h}}_k^{\mathrm{HWI}})^H
\mathbf{\Upsilon}_{\mathrm{BS}}
\widetilde{\mathbf{h}}_k^{\mathrm{HWI}}
+
\upsilon_{\mathrm{UE},k}
+
\sigma_k^2 .
\end{aligned}
\end{equation}

Consequently, the achievable data rate of UE $k$ under HWIs
can be expressed as
\[
R_k^{\mathrm{HWI}}
(\mathbf{v},\widetilde{\boldsymbol{\Phi}})
=
\log_2
\!\left(
1+\gamma_k^{\mathrm{HWI}}
\right).
\]
\subsection{Ideal-Hardware Special Case}
\label{subsec:ideal_special_case}

For benchmarking purposes, the ideal-hardware model is obtained by setting $
\delta_n^i=0,
\kappa_{\mathrm{BS}}=0,$ and $
\kappa_{\mathrm{UE}}=0,
\forall n\in\mathcal N,\; i\in\{t,r\}.$
Accordingly, $\widetilde{\boldsymbol{\Phi}}_i=\boldsymbol{\Phi}_i$, while the BS and UE distortion terms vanish. The channel at UE $k$, $k\in\mathcal{K}_i$, therefore reduces to
\(
\bar{\mathbf h}_k^H
=
\mathbf g_k^H
+
\mathbf h_k^H\boldsymbol{\Phi}_i\mathbf G.
\)
Accordingly, the received signal at UE $k$ is expressed as
\begin{align}
y_k
&=
(
\mathbf{g}_k^{H}
+
\mathbf{h}_k^{H}\boldsymbol{\Phi}_i\mathbf{G}
)
\sum_{l\in\mathcal{K}}
\mathbf{v}_ls_l
+
(
\mathbf{g}_k^{H}
+
\mathbf{h}_k^{H}\boldsymbol{\Phi}_i\mathbf{G}
)
\mathbf{v}_0s_0
+n_k.
\end{align}

By treating the sensing waveform as interference at the communication receivers, the SINR at UE $k$ is formulated as
\begin{equation}
\gamma_k
=
\frac{
\left|
\bar{\mathbf{h}}_k^{H}\mathbf{v}_k
\right|^2
}{
\sum\limits_{l\in\mathcal{K}\setminus\{k\}}
\left|
\bar{\mathbf{h}}_k^{H}\mathbf{v}_l
\right|^2
+
\left|
\bar{\mathbf{h}}_k^{H}\mathbf{v}_0
\right|^2
+
\sigma_k^2
}.
\end{equation}
Then, the achievable rate of UE $k$ is expressed as
\begin{align}
    R_k(\{\mathbf{v}_l\},\boldsymbol{\Phi})
=
\log_2(1+\gamma_k).
\end{align}


\section{Problem Statement and Formulation}
\label{sec3_problem}

Based on the unified system model in Section~\ref{sec2}, we consider the practical HWI-aware design, while retaining its ideal-hardware counterpart as a benchmark. This comparison characterizes the ideal-hardware performance limit and quantifies the degradation caused by STAR-RIS phase perturbations and signal-dependent transceiver distortion. Thus, the HWI-aware problem is presented first, followed by the ideal-hardware benchmark. Both formulations share the BS transmit-power, STAR-RIS ES, and sensing constraints, given by
\begin{subequations}
\label{prob:common_constraints}
\begin{align}
&
\sum_{k\in\mathcal{K}}
\|\mathbf{v}_k\|^2
+
\|\mathbf{v}_0\|^2
\le P_{\max},
\label{prob:wsr_main_power} \hspace{-3pt}
\\
&p(\theta)\ge\Gamma_{\mathrm{tar}}, \hspace{-3pt}
\label{prob:wsr_main_sensing}
\\
\label{eq:es_const}
& \rho_n^t,\rho_n^r\in[0,1], \rho_n^t+\rho_n^r=1,  \theta_n^t,\theta_n^r\in[0,2\pi), \forall n\in\mathcal N. \hspace{-3pt}
\end{align}
\end{subequations}
Constraint \eqref{prob:wsr_main_power} is the BS transmit-power budget, where $P_{\max}$ is the maximum transmit power. Constraint \eqref{prob:wsr_main_sensing} enforces the minimum sensing beampattern gain in the target direction, while constraints \eqref{eq:es_const} characterize the ES feasibility conditions of the STAR-RIS elements.

\subsection{HWI-Aware WSR Maximization}
\label{sec3_hwi_problem}

Based on the HWI-aware received-signal model above, the practical WSR maximization problem is formulated as
\begin{subequations}\label{prob:wsr_hwi_main}
\begin{align}
\max_{\mathbf v,\widetilde{\boldsymbol{\Phi}}_i,\widetilde{\boldsymbol{\Phi}}_r}
\quad
&
\sum\nolimits_{k\in\mathcal K}
\eta_k
R_k^{\mathrm{HWI}}
(\mathbf v,\widetilde{\boldsymbol{\Phi}})
\label{prob:wsr_hwi_main_obj}
\\
\text{s.t.}\quad
&
\eqref{prob:wsr_main_power}
-
\eqref{eq:es_const},
\label{prob:wsr_hwi_main_constraints}
\\
&
R_k^{\mathrm{HWI}}
(\mathbf v,\widetilde{\boldsymbol{\Phi}})
\ge
r_{\mathrm{qos}},
\quad
\forall k\in\mathcal K,
\label{prob:wsr_hwi_main_qos}
\end{align}
\end{subequations}
where \eqref{prob:wsr_hwi_main_qos} ensures the QoS requirement $r_{\mathrm{qos}}$ for each UE.

Problem
\eqref{prob:wsr_hwi_main}
is quite challenging due to the signal-dependent distortion terms. Specifically, both
$\mathbf\Upsilon_{\mathrm{BS}}$
and
$\upsilon_{\mathrm{UE},k}$
rely on the transmit covariance matrix and the effective cascaded channels, coupling the active beamforming vectors and passive STAR-RIS coefficients within the SINR expressions. Consequently, both \eqref{prob:wsr_hwi_main_obj} and \eqref{prob:wsr_hwi_main_qos} remain highly non-convex.

\subsection{Ideal-Hardware Benchmark Problem}
\label{subsec:ideal_problem}
For benchmark purposes, we consider the ideal-hardware specialization in Section~\ref{subsec:ideal_special_case}. Mathematically speaking, this optimization problem is formulated as
\begin{subequations}\label{prob:wsr_main}
\begin{align}
\max_{\{\mathbf{v}_k\},\,\mathbf{v}_0,\,\boldsymbol{\Phi}_t,\,\boldsymbol{\Phi}_r}
\quad
& \sum_{k\in\mathcal{K}}
\eta_k
R_k(\{\mathbf{v}_l\},\boldsymbol{\Phi}_i)
\label{prob:wsr_main_obj}
\\
\mathrm{s.t.}\quad
&
\eqref{prob:wsr_main_power}
-
\eqref{eq:es_const},
\\
&
R_k(\{\mathbf{v}_l\},\boldsymbol{\Phi}_i)
\ge
r_{\mathrm{qos}},
\quad
\forall k\in\mathcal{K}.
\label{prob:wsr_main_qos}
\end{align}
\end{subequations}


Problem \eqref{prob:wsr_main} is non-convex due to three coupled structures. First, the fractional SINR expressions bilinearly couple the active beamforming vectors and STAR-RIS coefficients through the cascaded channels, rendering the objective \eqref{prob:wsr_main_obj} and QoS constraints \eqref{prob:wsr_main_qos} non-convex. Second, the sensing constraint \eqref{prob:wsr_main_sensing} imposes a reverse-convex quadratic inequality on the transmit covariance matrix. Third, the ES constraints \eqref{eq:es_const} couple the STAR-RIS amplitudes and impose non-convex phase constraints. These structures preclude direct optimization of Problem \eqref{prob:wsr_main}.

\section{Proposed AO-based SCA Framework}
\label{sec4_framework}

The HWI-aware and ideal-hardware problems formulated in Section~\ref{sec3_problem} share the coupling between the active BS beamformers and passive STAR-RIS coefficients. We therefore develop an AO-based SCA framework by decomposing each original problem into two sub-optimization problems: \textbf{1)} active beamforming
and \textbf{2)} passive STAR-RIS optimizations. In particular, the HWI-aware solution for Problem~\ref{prob:wsr_hwi_main} is presented first, followed by the ideal-hardware benchmark design of Problem~\eqref{prob:wsr_main}. To facilitate the analyses, we define the STAR-RIS phase matrix 
\(\boldsymbol{\Phi}_i
=
\operatorname{diag}(\mathbf q_i)\), with the coefficient vectors
\[
\mathbf q_i
\triangleq
\big[
\sqrt{\rho_1^i}e^{j\theta_1^i},
\dots,
\sqrt{\rho_N^i}e^{j\theta_N^i}
\big]^T,
\quad
i\in\{t,r\}.
\]

\subsection{HWI-Aware AO-Based SCA Design}
\label{sec3_hwi}

\subsubsection{Beamformer Optimization under Hardware Impairments}
\label{subsec:BF_AOSCA_HWI}

The signal-dependent distortion powers at the BS and the $k$-th UE are given by
\begin{align}
\label{eq:distortion_hwi}
    \begin{cases}
        \mathbf{\Upsilon}_{\mathrm{BS}}
=
\kappa_{\mathrm{BS}}^2\,\mathrm{diag}(\mathbf W),
\\
\upsilon_{\mathrm{UE},k}
=
\kappa_{\mathrm{UE}}^2
(\widetilde{\mathbf h}_k^{\mathrm{HWI}})^H
\mathbf W
\widetilde{\mathbf h}_k^{\mathrm{HWI}}.
    \end{cases}
\end{align}

To normalize the thermal noise, let
\(
\bar{\mathbf h}_k^H
\triangleq
(\widetilde{\mathbf h}_k^{\mathrm{HWI}})^H/\sigma_k
\). Then, the desired signal term is written as
$z_k(\mathbf v)
\triangleq
\bar{\mathbf h}_k^H\mathbf v_k$.
The interference-plus-distortion-and-noise term is given by $\psi_k^{\mathrm{HWI}}(\mathbf v) \triangleq
\sum_{\ell\neq k}|\bar{\mathbf h}_k^H\mathbf v_\ell|^2
+|\bar{\mathbf h}_k^H\mathbf v_0|^2
+1
+\bar{\mathbf h}_k^H\boldsymbol{\Upsilon}_{\mathrm{BS}}\bar{\mathbf h}_k
+\kappa_{\mathrm{UE}}^2\bar{\mathbf h}_k^H\mathbf W\bar{\mathbf h}_k$.
Accordingly, the HWI-aware SINR is given by
\[
\gamma_k^{\mathrm{HWI}}(\mathbf v)
=
{
|z_k(\mathbf v)|^2
}/{
\psi_k^{\mathrm{HWI}}(\mathbf v)
}.
\]

To convexify the BS distortion term, an auxiliary vector
$\mathbf p\succeq\mathbf0$
is added to upper-bound the diagonal entries of
$\mathbf W$:
\begin{equation}
\label{eq:p_def_hwi_bf}
p_m
\ge
\sum_{\ell\in\mathcal K}
|[\mathbf v_\ell]_m|^2
+
|[\mathbf v_0]_m|^2,
\quad
\forall m,
\end{equation}
which yields
\begin{equation}\label{eq:BSdist_bound_hwi_bf}
\begin{aligned}
\bar{\mathbf h}_k^H
\mathbf\Upsilon_{\mathrm{BS}}
\bar{\mathbf h}_k
&=
\kappa_{\mathrm{BS}}^2
\bar{\mathbf h}_k^H
\operatorname{diag}(\mathbf W)
\bar{\mathbf h}_k
\le
\kappa_{\mathrm{BS}}^2
\sum_{m=1}^{M_b}
|[\bar{\mathbf h}_k]_m|^2
p_m .
\end{aligned}
\end{equation}

To construct an SCA surrogate for the achievable rate, we first define $
\bar z_k
\triangleq
z_k(\mathbf v^{(t-1)}),
\bar\psi_k
\triangleq
\psi_k^{\mathrm{HWI}}(\mathbf v^{(t-1)}),$ and $
\alpha_k
\triangleq
\frac{
|\bar z_k|^2
}{
\bar\psi_k+|\bar z_k|^2
}$. Then, using the first-order approximation, the following concave surrogate lower bound is obtained:
\begin{equation}\label{eq:mk_bf_hwi}
\begin{aligned}
\widehat R_k^{\mathrm{HWI}}(\mathbf v)
&\ge
\log_2
\!\left(
1+\frac{|\bar z_k|^2}{\bar\psi_k}
\right)
\\
&\quad+
\frac{\alpha_k}{\ln2}
\left[
2
-
\frac{|z_k(\mathbf v)|^2}{|\bar z_k|^2}
-
\frac{\psi_k^{\mathrm{HWI}}(\mathbf v)}{\bar\psi_k}
\right].
\end{aligned}
\end{equation}

The proposed surrogate function is locally tight at
$\mathbf v^{(t-1)}$ and satisfies the first-order consistency condition. However, to further handle the sensing constraint, we define $u_\ell^{(t-1)}
=
\mathbf a^H(\theta_{\rm tar})
\mathbf v_\ell^{(t-1)},
$ and $
u_0^{(t-1)}
=
\mathbf a^H(\theta_{\rm tar})
\mathbf v_0^{(t-1)}$.
Thus, the corresponding affine lower bound is expressed as
\begin{equation}\label{prop:beam_lb_hwi}
\begin{aligned}
&
\sum_{\ell\in\mathcal K}
(
2\Re
\{
(u_\ell^{(t-1)})^*
\mathbf a^H(\theta_{\rm tar})
\mathbf v_\ell
\}
-
|u_\ell^{(t-1)}|^2
)
\\
&+
(
2\Re
\{
(u_0^{(t-1)})^*
\mathbf a^H(\theta_{\rm tar})
\mathbf v_0
\}
-
|u_0^{(t-1)}|^2
)
\ge
\Gamma_{\rm tar}.
\end{aligned}
\end{equation}

Putting it all together, the SCA subproblem behaves
\begin{subequations}\label{eq:BF_SCA_HWI}
\begin{align}
\max_{\mathbf v,\mathbf p}
\quad
&
\sum_{k\in\mathcal K}
\eta_k
\widehat R_k^{\mathrm{HWI}}(\mathbf v)
\label{eq:BF_SCA_HWI_obj}
\\
\text{s.t.}\quad
&
\eqref{prob:wsr_main_power},\eqref{prop:beam_lb_hwi},
\eqref{eq:p_def_hwi_bf},
\bar\psi_k>0,
|\bar z_k|^2>0,
\;
\forall k\in\mathcal K,
\label{eq:BF_SCA_HWI_pos}
\\
&
\widehat R_k^{\mathrm{HWI}}(\mathbf v)
\ge
r_{\mathrm{qos}},
\quad
\forall k\in\mathcal K .
\end{align}
\end{subequations}

The surrogate objective function $\widehat R_k^{\mathrm{HWI}}(\mathbf v)$ is concave with respect to (w.r.t) the optimization variables, while \eqref{prop:beam_lb_hwi} and \eqref{eq:p_def_hwi_bf} define convex constraints. Accordingly, problem \eqref{eq:BF_SCA_HWI} is implemented in CVX and solved using the SDPT3 solver. The per-iteration complexity is dominated by the interior-point updates associated with the beamforming variables.

\subsubsection{STAR-RIS Coefficient Optimization under HWIs}
\label{subsec:Phi_AOSCA_HWI}

Given fixed BS beamformers
$\{\mathbf v_k\}_{k=1}^{K}$
and
$\mathbf v_0$,
the passive STAR-RIS coefficients are optimized.
Then,
we define
$\mathbf q_t,\mathbf q_r\in\mathbb C^{N\times1}$
as the transmission and reflection coefficient vectors,
with $\boldsymbol{\Phi}_i
=
\operatorname{diag}(\mathbf q_i),
i\in\{t,r\}$.
Under this representation, the ES constraints can be equivalently expressed as a set of equations
\begin{subequations}
\label{eq:es_constraints_q}
\begin{empheq}[left = {\empheqlbrace}]{align}
\label{eq:es_constraints_q-a}
&0 \le |[\mathbf q_t]_n|^2 \le 1,
\\
\label{eq:es_constraints_q-b}
&0 \le |[\mathbf q_r]_n|^2 \le 1, \quad \forall n\in\mathcal N,
\\
\label{eq:es_constraints_q-c}
&|[\mathbf q_t]_n|^2 + |[\mathbf q_r]_n|^2 = 1, \quad \forall n\in\mathcal N.
\end{empheq}
\end{subequations}

Using the identity
$\mathbf a^H\operatorname{diag}(\mathbf b)
=
\mathbf b^T\operatorname{diag}(\mathbf a^*)$,
the effective cascaded channel is rewritten as
\begin{equation}
\label{eq:Hk_q}
\widetilde{\mathbf h}_k^H(\mathbf q_k)
\triangleq
\mathbf g_k^H
+
\mathbf h_k^H
\operatorname{diag}(\mathbf q_k)
\mathbf G
\in
\mathbb C^{1\times M_b}.
\end{equation}
Accordingly, the desired signal term is expressed as
$z_k(\mathbf q_k)
\triangleq
\widetilde{\mathbf h}_k^H(\mathbf q_k)\mathbf v_k$,
while the multiuser interference and radar leakage terms are
given, respectively, by
\begin{equation}
\iota_k(\mathbf q_k)
\triangleq
\sum_{\ell\neq k}
|
\widetilde{\mathbf h}_k^H(\mathbf q_k)\mathbf v_\ell
|^2,
\,
r_k(\mathbf q_k)
\triangleq
|
\widetilde{\mathbf h}_k^H(\mathbf q_k)\mathbf v_0
|^2.
\end{equation}

To relax HWIs's issues, let
$\mathbf p
\triangleq
\operatorname{diag}(\mathbf W)
\in
\mathbb R_+^{M_b\times1}$.
Then, the received BS distortion power at UE $k$ becomes
\begin{equation}
\label{eq:bs_dist_q}
\mathcal I_k^{\mathrm{BS}}(\mathbf q_k)
=
\kappa_{\mathrm{BS}}^2
\sum_{m=1}^{M_b}
p_m
|
[\widetilde{\mathbf h}_k^H(\mathbf q_k)]_m
|^2,
\end{equation}
while the UE distortion term is expressed as
\begin{equation}
\label{eq:ue_dist_q}
\mathcal I_k^{\mathrm{UE}}(\mathbf q_k)
=
\kappa_{\mathrm{UE}}^2
\left[
\sum_{\ell=1}^{K}
|
\widetilde{\mathbf h}_k^H(\mathbf q_k)\mathbf v_\ell
|^2
+
|
\widetilde{\mathbf h}_k^H(\mathbf q_k)\mathbf v_0
|^2
\right].
\end{equation}

These allow us to obtain the resultant HWI-aware SINR as
\begin{equation}
\label{eq:SINR_q_hwi}
\gamma_k(\mathbf q_k)
=
{
|z_k(\mathbf q_k)|^2
}
/{\psi_k^{\mathrm{HWI}}(\mathbf q_k)},
\end{equation}
where
$\psi_k^{\mathrm{HWI}}(\mathbf q_k)
\triangleq
\iota_k(\mathbf q_k)
+
r_k(\mathbf q_k)
+
\sigma^2
+
\mathcal I_k^{\mathrm{BS}}(\mathbf q_k)
+
\mathcal I_k^{\mathrm{UE}}(\mathbf q_k)$ and the corresponding WSR is written as
\begin{equation}
\label{eq:WSR_q_hwi}
\mathrm{WSR}(\mathbf q_k)
=
\sum_{k=1}^{K}
\eta_k
\log_2
\left(
1+\gamma_k(\mathbf q_k)
\right).
\end{equation}

Thus, the STAR-RIS optimization subproblem becomes
\begin{subequations}
\label{prob:sub2_hwi}
\begin{align}
\max_{\mathbf q_t,\mathbf q_r}
\quad
&
\sum_{k=1}^{K}
\eta_k
\log_2
\left(
1+\gamma_k(\mathbf q_k)
\right)
\\
\text{s.t.}\quad
&
\eqref{eq:es_constraints_q},
\quad
R_k^{\mathrm{HWI}}(\mathbf q_k)
\ge
r_{\mathrm{qos}},
\quad
\forall k\in\mathcal K.
\end{align}
\end{subequations}

\begingroup

The equality ES constraint in \eqref{eq:es_constraints_q}
is non-convex w.r.t the STAR-RIS coefficients.
Instead of employing a post-optimization projection, we first relax it as
\begin{align}
\label{eq:es_constraints_q-equal}
    |[\mathbf q_t]_n|^2 + |[\mathbf q_r]_n|^2 \leq 1, \quad \forall n\in\mathcal N,
\end{align}
and then incorporate the equality requirement into the objective
through a penalty function. To proceed, we define
\begin{align}
    S(\mathbf q_t,\mathbf q_r)
\triangleq
\sum_{n=1}^{N}
\left(
|[\mathbf q_t]_n|^2
+
|[\mathbf q_r]_n|^2
\right).
\label{S-condition-qt-qr}
\end{align}
Under the above relaxed ES constraints,
$S(\mathbf q_t,\mathbf q_r)\leq N$.
We therefore introduce the ES penalty function
\begin{align}
    f_{\mathrm{ES}}(\mathbf q_t,\mathbf q_r)
\triangleq
{1}/{N}
-
{1}/{
S(\mathbf q_t,\mathbf q_r)
}.
\label{f-ES-condition-qt-qr}
\end{align}
It follows that $f_{\mathrm{ES}}(\mathbf q_t,\mathbf q_r)
\leq 0$ and $f_{\mathrm{ES}}(\mathbf q_t,\mathbf q_r)=0$
holds if and only if \eqref{eq:es_constraints_q-c} remains valid.

Since $S(\mathbf q_t,\mathbf q_r)$ is convex, we construct
an affine global lower bound at the current SCA point
$\{\mathbf q_t^{(i-1)},\mathbf q_r^{(i-1)}\}$. To do so, we introduce the following lemma.

\begin{lemma}[Affine lower bound for a quadratic function]
\label{lem:quadratic_lb}
For $N_y$ local point $\mathbf{x}^{(t-1)}\in\mathbb{C}^{L}$, the quadratic function $\|\mathbf{x}\|^{2}$ satisfies
\begin{equation}
\|\mathbf{x}\|^{2}
\ge
2\Re\!\left\{
(\mathbf{x}^{(t-1)})^{H}\mathbf{x}
\right\}
-
\|\mathbf{x}^{(t-1)}\|^{2}.
\end{equation}

\begin{proof}
Let $\mathbf{x}_0=\mathbf{x}^{(t-1)}$. Since
$\|\mathbf{x}-\mathbf{x}_0\|^2\ge0$, we have
\[
\|\mathbf{x}\|^2
-
2\Re\!\left\{
\mathbf{x}_0^H\mathbf{x}
\right\}
+
\|\mathbf{x}_0\|^2
\ge0,
\]
where the equality follows from expanding
$\|\mathbf{x}-\mathbf{x}_0\|^2$.
Rearranging the above inequality and substituting
$\mathbf{x}_0=\mathbf{x}^{(t-1)}$
complete the proof.
\end{proof}
\end{lemma}


In Lemma~\ref{lem:quadratic_lb}, we define
\[
\begin{aligned}
\underline S^{(i)}(\mathbf q_t,\mathbf q_r)
\triangleq
&\sum_{n=1}^{N}
\Big(
2\Re\big\{
([\mathbf q_t^{(i-1)}]_n)^*
[\mathbf q_t]_n
\big\}
-
|[\mathbf q_t^{(i-1)}]_n|^2
\\
&+
2\Re\big\{
([\mathbf q_r^{(i-1)}]_n)^*
[\mathbf q_r]_n
\big\}
-
|[\mathbf q_r^{(i-1)}]_n|^2
\Big).
\end{aligned}
\]
So, a concave lower-bound surrogate of the ES
penalty writes
\[
\widehat f_{\mathrm{ES}}^{(i)}
(\mathbf q_t,\mathbf q_r)
\triangleq
{1}/{N}
-
{1}/{
\underline S^{(i)}(\mathbf q_t,\mathbf q_r)
}.
\]
Since \(
S(\mathbf q_t,\mathbf q_r)
\geq
\underline S^{(i)}(\mathbf q_t,\mathbf q_r),\)
we have
\[
f_{\mathrm{ES}}(\mathbf q_t,\mathbf q_r)
\geq
\widehat f_{\mathrm{ES}}^{(i)}
(\mathbf q_t,\mathbf q_r),
\]
where equality holds at
$\{\mathbf q_t^{(i-1)},\mathbf q_r^{(i-1)}\}$.

\endgroup

Furthermore, the fractional SINR structure renders problem
\eqref{prob:sub2_hwi}
non-convex.
To handle this difficulty, SCA is employed to construct a
concave surrogate lower bound.

At iteration $i$, let us define $
\bar z_k
\triangleq
z_k(\mathbf q_k^{(i-1)}),
$ and $
\bar\psi_k
\triangleq
\psi_k^{\mathrm{HWI}}
(\mathbf q_k^{(i-1)})$.
The corresponding constants are given by
$s_k^{(i-1)}
=
|\bar z_k|^2/\bar\psi_k$,
$c_{1k}
=
\log_2(1+s_k^{(i-1)})$, $
c_{2k}
=
{
{
s_k^{(i-1)}
}{/[
(1+s_k^{(i-1)})\ln 2
]}},$
and
$c_{3k}=|\bar z_k|^2$.
The linearized desired signal term is expressed as
\[
\widetilde z_k(\mathbf q_k)
\triangleq
2\Re
\left\{
\bar z_k^*
z_k(\mathbf q_k)
\right\}
-
c_{3k}.
\]
Applying the first-order approximation yields the following
concave surrogate lower bound:
\begin{equation}
\label{eq:mk_sub2}
\widehat R_k(\mathbf q_k)
\ge
c_{1k}
+
c_{2k}
\left(
2
-
\frac{c_{3k}}
{\widetilde z_k(\mathbf q_k)}
-
\frac{
\psi_k^{\mathrm{HWI}}(\mathbf q_k)
}{
\bar\psi_k
}
\right).
\end{equation}

The surrogate function is locally tight at
$\mathbf q_k^{(i-1)}$
and satisfies the first-order optimal condition.
The SCA subproblem is thus formulated as follows:
\begin{subequations}
\label{prob:sub2_sca_hwi}
\begin{align}
\max_{\mathbf q_t,\mathbf q_r,t}
\quad
&
{
t
+
\mu
\widehat f_{\mathrm{ES}}^{(i)}
(\mathbf q_t,\mathbf q_r)}
\label{prob:sub2_sca_hwi_obj}
\\
\text{s.t.}\quad
&
\sum_{k=1}^{K}
\eta_k
\widehat R_k(\mathbf q_k)
\geq t,
\quad
{
\widetilde z_k(\mathbf q_k)
\geq
\epsilon},
\quad
\forall k,
\label{prob:sub2_sca_hwi_rate}
\\
&
\widehat R_k(\mathbf q_k)
\geq
r_{\mathrm{qos}},
\quad
\forall k\in\mathcal K,
\label{prob:sub2_sca_hwi_qos}
\\
&
\eqref{eq:es_constraints_q-equal},
\quad
{
\underline S^{(i)}
(\mathbf q_t,\mathbf q_r)
\geq
\epsilon_{\mathrm{ES}},}
\label{prob:sub2_sca_hwi_penalty}
\end{align}
\end{subequations}
{
where $\mu>0$ denotes the ES penalty coefficient,
$\epsilon>0$ guarantees the validity of the rate surrogate,
and $\epsilon_{\mathrm{ES}}>0$ ensures that the reciprocal
penalty surrogate is well defined.}
\begingroup

\begin{proposition}
\label{prop:penalty_ris_hwi}
For a fixed penalty coefficient $\mu>0$ and a given SCA
operating point
$\{\mathbf q_t^{(i-1)},\mathbf q_r^{(i-1)}\}$,
Problem~\eqref{prob:sub2_sca_hwi} is a convex optimization
problem. Moreover, under the relaxed ES constraint
\eqref{eq:es_constraints_q-equal}, the penalty function
$f_{\mathrm{ES}}(\mathbf q_t,\mathbf q_r)$ is non-positive
and vanishes if and only if the original element-wise ES
equality constraints in \eqref{eq:es_constraints_q} are
satisfied.
\end{proposition}

\begin{proof}
Each STAR-RIS element satisfies \eqref{eq:es_constraints_q-equal}.
Consequently, $
S(\mathbf q_t,\mathbf q_r)
\leq N$, and the equality hold if
$f_{\mathrm{ES}}(\mathbf q_t,\mathbf q_r)=0$.
Since every term of $S(\mathbf q_t,\mathbf q_r)$ is upper
bounded by one, $S(\mathbf q_t,\mathbf q_r)=N$ holds if and
only if \eqref{eq:es_constraints_q-c} is valid. Therefore, a vanishing ES penalty satisfies the
original constraints \eqref{eq:es_constraints_q-c}.

We next establish the convexity of
Problem~\eqref{prob:sub2_sca_hwi}.
The function
$\underline S^{(i)}(\mathbf q_t,\mathbf q_r)$
is affine in $\mathbf q_t$ and $\mathbf q_r$.
Since $-1/x$ is concave for $x>0$,
$\widehat f_{\mathrm{ES}}^{(i)}
(\mathbf q_t,\mathbf q_r)$
is concave over the domain specified by $\underline S^{(i)}
(\mathbf q_t,\mathbf q_r)
\geq
\epsilon_{\mathrm{ES}}$. To be specific, by construction,
$\widehat R_k(\mathbf q_k)$
is concave over
$\widetilde z_k(\mathbf q_k)>0$.
Therefore, $t+
\mu
\widehat f_{\mathrm{ES}}^{(i)}
(\mathbf q_t,\mathbf q_r)$
is a concave objective function.
Constraints in
\eqref{prob:sub2_sca_hwi_rate}
and
\eqref{prob:sub2_sca_hwi_qos}
are superlevel sets of concave functions, while
\eqref{eq:es_constraints_q-equal}
is convex because its left-hand side is a convex quadratic
function, and the second term of \eqref{prob:sub2_sca_hwi_penalty}
is affine.
Hence,
Problem~\eqref{prob:sub2_sca_hwi}
maximizes a concave objective over a convex feasible set and
is thus a convex optimization problem.
\end{proof}

\endgroup
\begingroup

{
For notational convenience, we further define the corresponding weight-sum rate and the penalty function
\begin{align}
    F(\mathbf v,\mathbf q_t,\mathbf q_r)
&\triangleq
\sum_{k\in\mathcal K}
\eta_k
R_k^{\mathrm{HWI}}
(\mathbf v,\mathbf q_k),\\
\mathcal J_{\mu}
(\mathbf v,\mathbf q_t,\mathbf q_r)
&\triangleq
F(\mathbf v,\mathbf q_t,\mathbf q_r)
+
\mu
f_{\mathrm{ES}}(\mathbf q_t,\mathbf q_r).
\end{align}

}
\begin{algorithm}[t]

\caption{{Proposed AO-based SCA Algorithm for HWI-Aware ISAC Design}}
\label{alg:AO_SCA_HWI}

{
\begin{algorithmic}[1]

\State \textbf{Input:}
$\mathbf G$, $\{\mathbf h_k\}$, $\{\mathbf g_k\}$, $\kappa_{\mathrm{BS}}$ $\kappa_{\mathrm{UE}}$,
$P_{\max}$, $\Gamma_{\mathrm{tar}}$,
 $r_{\mathrm{qos}}$,
 $\epsilon$,
 $\mu>0$, and 
$\varepsilon_{\mathrm{ES}}>0$.

\State \textbf{Initialize:}
Feasible points $\{\mathbf v_k^{(0)}\}$,
$\mathbf v_0^{(0)}$, and
$\{\mathbf q_t^{(0)},\mathbf q_r^{(0)}\}$.
Set iteration index $i=0$.

\State Compute the initial penalized objective $\mathcal J_{\mu}^{(0)}$.

\Repeat

\State For fixed
$\{\mathbf q_t^{(i)},\mathbf q_r^{(i)}\}$,
update $\{\mathbf v_k^{(i+1)}\}$ and
$\mathbf v_0^{(i+1)}$
by solving Problem~\eqref{eq:BF_SCA_HWI}.

\State For fixed
$\{\mathbf v_k^{(i+1)}\}$ and
$\mathbf v_0^{(i+1)}$,
update
$\{\mathbf q_t^{(i+1)},\mathbf q_r^{(i+1)}\}$
by solving the penalized SCA subproblem
\eqref{prob:sub2_sca_hwi}.

\State Evaluate the ES equality residual as
\[
e_{\mathrm{ES}}^{(i+1)}
=
\max_{n\in\mathcal N}
|
|[\mathbf q_t^{(i+1)}]_n|^2
+
|[\mathbf q_r^{(i+1)}]_n|^2
-
1
|.
\]

\State Evaluate the penalized objective $\mathcal J_{\mu}^{(i+1)}$.

\State Compute the relative objective variation
\[
\Delta^{(i+1)}
=
{
|
\mathcal J_{\mu}^{(i+1)}
-
\mathcal J_{\mu}^{(i)}
|
}/{
\max\{
1,
|
\mathcal J_{\mu}^{(i)}
|
\}
}.
\]

\State Set $i\leftarrow i+1$.

\Until{
$\Delta^{(i)}\leq\epsilon$
and
$e_{\mathrm{ES}}^{(i)}
\leq
\varepsilon_{\mathrm{ES}}$
}

\State \textbf{Output:}
Optimized beamformers
$\{\mathbf v_k^\star\}$,
$\mathbf v_0^\star$,
and STAR-RIS coefficients
$\{\mathbf q_t^\star,\mathbf q_r^\star\}$.

\end{algorithmic}
}
\end{algorithm}

\endgroup

\subsubsection{Algorithm, Convergence and Complexity Analysis under HWIs}
\label{subsec:convergence_complexity_hwi}
{
This subsection first describes the proposed
HWI-aware AO-based SCA algorithm and then analyzes its convergence
behavior and computational complexity.

\paragraph{AO-Based Iterative Procedure}


The proposed HWI-aware AO-based SCA procedure is summarized in Algorithm~\ref{alg:AO_SCA_HWI}. To implement, it is required to first initialize the STAR-RIS coefficients satisfying
\eqref{eq:es_const}, and then jointly determine $(\{\mathbf v_k\},\,\mathbf v_0,\,\alpha)$ that maximizes a non-negative minimum normalized feasibility margin $\alpha$ subject to the constraints
\begin{align}
    \eqref{prob:wsr_main_power},
\frac{p(\theta)}{\Gamma_{\mathrm{tar}}}-1
\ge \alpha, \frac{
R_k^{\mathrm{HWI}}
(\mathbf v,\widetilde{\boldsymbol{\Phi}}_i^{(0)})
}{r_{\mathrm{qos}}}-1
\ge \alpha, \forall k\in\mathcal K,\nonumber
\end{align}
where 
$\widetilde{\boldsymbol{\Phi}}_i^{(0)}$ is the fixed matrices. Upon these feasible points, the algorithm alternately optimizes the active BS beamformers and passive STAR-RIS coefficients. At AO iteration $i$, the communication and sensing beamformers are updated by solving \eqref{eq:BF_SCA_HWI} for fixed ${\mathbf q_t^{(i)},\mathbf q_r^{(i)}}$, after which the STAR-RIS transmission and reflection coefficients are updated via the penalized SCA subproblem \eqref{prob:sub2_sca_hwi}. The penalty term directly promotes the original ES equality constraint, eliminating post-optimization normalization. The procedure evaluates the ES residual $e_{\mathrm{ES}}^{(i)}$ after each iteration and terminates when $\Delta^{(i)}\leq\epsilon$ and $e_{\mathrm{ES}}^{(i)}\leq\varepsilon_{\mathrm{ES}}$, where $\epsilon$ is the prescribed convergence tolerance. The resultant solution comprises the optimized communication and sensing beamformers and STAR-RIS coefficients.

\paragraph{Convergence Analysis}

\begingroup

Using the penalized objective $\mathcal J_\mu$ defined above, for fixed STAR-RIS coefficients
$\{\mathbf q_t^{(i)},\mathbf q_r^{(i)}\}$,
the ES penalty term remains constant during the active
beamforming update. Since the active-beamforming surrogate
is a locally tight lower bound satisfying the first-order
consistency condition, the solution
$\mathbf v^{(i+1)}$ satisfies
\begin{equation}
\label{eq:conv_active_penalty}
\mathcal J_{\mu}
(
\mathbf v^{(i+1)},
\mathbf q_t^{(i)},
\mathbf q_r^{(i)}
)
\geq
\mathcal J_{\mu}
(
\mathbf v^{(i)},
\mathbf q_t^{(i)},
\mathbf q_r^{(i)}
).
\end{equation}

For fixed $\mathbf v^{(i+1)}$, $\widehat R_k(\mathbf q_k)$ and $\widehat f_{\mathrm{ES}}^{(i)}
(\mathbf q_t,\mathbf q_r)$ are respectively the the global lower bounds of
$R_k^{\mathrm{HWI}}(\mathbf q_k)$ and
$f_{\mathrm{ES}}(\mathbf q_t,\mathbf q_r)$.
Therefore, the passive SCA
surrogate objective holds
\begin{align}
\label{eq:conv_lower_bound_penalty}
\widehat{\mathcal J}_{q}^{(i)}
(\mathbf q_t,\mathbf q_r)
&\triangleq
\sum_{k\in\mathcal K}
\eta_k
\widehat R_k(\mathbf q_k)
+
\mu
\widehat f_{\mathrm{ES}}^{(i)}
(\mathbf q_t,\mathbf q_r)\nonumber\\
&\leq
\mathcal J_{\mu}
(
\mathbf v^{(i+1)},
\mathbf q_t,\mathbf q_r
).
\end{align}

Furthermore, both surrogate functions are locally tight at
the current SCA point, yielding
\begin{equation}
\label{eq:conv_tight_penalty}
\widehat{\mathcal J}_{q}^{(i)}
(
\mathbf q_t^{(i)},
\mathbf q_r^{(i)}
)
=
\mathcal J_{\mu}
(
\mathbf v^{(i+1)},
\mathbf q_t^{(i)},
\mathbf q_r^{(i)}
).
\end{equation}

Since
$\{\mathbf q_t^{(i+1)},\mathbf q_r^{(i+1)}\}$
is obtained by solving
\eqref{prob:sub2_sca_hwi},
we have
\begin{align}
&
\widehat{\mathcal J}_{q}^{(i)}
(
\mathbf q_t^{(i+1)},
\mathbf q_r^{(i+1)}
)\geq
\widehat{\mathcal J}_{q}^{(i)}
(
\mathbf q_t^{(i)},
\mathbf q_r^{(i)}
).
\label{eq:conv_surrogate_penalty}
\end{align}

Combining
\eqref{eq:conv_lower_bound_penalty}--%
\eqref{eq:conv_surrogate_penalty}
gives
\begin{align}
&
\mathcal J_{\mu}
(
\mathbf v^{(i+1)},
\mathbf q_t^{(i+1)},
\mathbf q_r^{(i+1)}
)\geq
\mathcal J_{\mu}
(
\mathbf v^{(i+1)},
\mathbf q_t^{(i)},
\mathbf q_r^{(i)}
).
\label{eq:conv_passive_penalty}
\end{align}

Together with \eqref{eq:conv_active_penalty}, this yields
\begin{equation}
\label{eq:conv_total_penalty}
\mathcal J_{\mu}^{(i+1)}
\geq
\mathcal J_{\mu}^{(i)}.
\end{equation}
Hence, the proposed penalty-based AO-SCA procedure generates
a non-decreasing sequence of penalized objective values.

Since the transmit power is bounded by $P_{\max}$ and the thermal-noise power is strictly positive, the achievable WSR is upper bounded. Furthermore, $f_{\mathrm{ES}}(\mathbf q_t,\mathbf q_r)\leq0$ over the relaxed ES feasible set, ensuring that $\mathcal J_{\mu}$ is upper bounded and ${\mathcal J_{\mu}^{(i)}}$ converges to a finite value. With a sufficiently large $\mu$, the converged solution satisfies
\(
e_{\mathrm{ES}}
\triangleq
\max_{n\in\mathcal N}
|
|[\mathbf q_t]_n|^2
+
|[\mathbf q_r]_n|^2
-
1|\leq\varepsilon_{\mathrm{ES}},
\)
thereby satisfying the original ES equality constraint within the prescribed numerical accuracy.




\endgroup

\paragraph{Computational Complexity Analysis}
Following \cite{8359418}, an interior-point method for a convex problem with $n$ real-valued variables and $m$ convex constraints has complexity $\mathcal O(n^2m^{2.5}+m^{3.5})$. Based on this foundation, the active and passive subproblems involve approximately $n_1=2M_b(K+1)+M_b$ and $n_2=4N$ real-valued variables, respectively, with corresponding constraint counts $m_1$ and $m_2$. Therefore, over $I_{\mathrm{AO}}$ AO iterations, the overall complexity is $\mathcal O\!\left[I_{\mathrm{AO}}\left(n_1^2m_1^{2.5}+m_1^{3.5}+n_2^2m_2^{2.5}+m_2^{3.5}\right)\right]$, confirming the polynomial complexity of the proposed framework.

}


\subsection{Ideal-Hardware Benchmark Design}\label{subsec:ideal_solution}

\subsubsection{Active Beamforming Optimization}
In this optimization block, the STAR-RIS coefficients $\boldsymbol{\Phi}_i$ are fixed, while the communication beamformers $\{\mathbf{v}_k\}_{k=1}^{K}$ and sensing beamformer $\mathbf{v}_0$ are optimized. The SINR of UE $k$ is expressed as
\begin{equation}\label{eq:active_sinr}
\gamma_k
= {|z_k|^2}- {\psi_k},
\end{equation}
where
\(
z_k
\triangleq
\widetilde{\mathbf{h}}_k^{H}\mathbf{v}_k
\) and
$\psi_k
\triangleq
\sum_{\ell\neq k}
|\widetilde{\mathbf{h}}_k^{H}\mathbf{v}_{\ell}|^2
+
|\widetilde{\mathbf{h}}_k^{H}\mathbf{v}_0|^2
+
\sigma_k^2$.

Since the achievable rates depend on coupled SINR expressions, the WSR objective is non-concave w.r.t the beamforming variables. Hence, SCA is employed to construct a concave lower-bound approximation of the achievable rate at iteration $t$, denoted by $\widehat{R}_k(\mathbf v)$. Also, the sensing constraint in \eqref{prob:wsr_main_sensing} is non-convex since it involves a reverse-convex quadratic inequality. 

Define
\(
\bar u_k
\triangleq
\mathbf a^H(\theta)\mathbf v_k^{(t-1)}\) and \(
\bar u_0
\triangleq
\mathbf a^H(\theta)\mathbf v_0^{(t-1)}.
\)
Applying Lemma~\ref{lem:quadratic_lb}, the constraint \eqref{prob:wsr_main_sensing} can be rewritten as
\begin{align}
\eqref{prob:wsr_main_sensing}
\Leftrightarrow &\sum_{k\in\mathcal K}
|\mathbf a^H(\theta)\mathbf v_k|^2
+
|\mathbf a^H(\theta)\mathbf v_0|^2
\ge
\Gamma_{\mathrm{tar}} \nonumber\\
\label{eq:sensing_affine_lb}
\Rightarrow &
\sum_{k\in\mathcal K}
\left(
2\Re\!\left\{
\bar u_k^*
\mathbf a^H(\theta)\mathbf v_k
\right\}
-
|\bar u_k|^2
\right)
\nonumber\\
&\quad+
\left(
2\Re\!\left\{
\bar u_0^*
\mathbf a^H(\theta)\mathbf v_0
\right\}
-
|\bar u_0|^2
\right)
\ge
\Gamma_{\mathrm{tar}}.
\end{align}

Upon the above approaches, the resultant SCA subproblem can be reformulated as
\begin{subequations}\label{prob:active_sca}
\begin{align}
\max_{\mathbf v}
\quad
&
\sum_{k\in\mathcal K}
\eta_k
\widehat R_k(\mathbf v)
\\
\text{s.t.}
\quad
&
\eqref{prob:wsr_main_power},
\quad
\eqref{eq:sensing_affine_lb},
\\
&
2\Re\{
(z_k^{(t-1)})^* z_k
\}
-
|z_k^{(t-1)}|^2
\ge
0,
\quad
\forall k\in\mathcal K,
\\
&
\widehat R_k(\mathbf v)
\ge
r_{\mathrm{qos}},
\quad
\forall k\in\mathcal K.
\end{align}
\end{subequations}

Problem \eqref{prob:active_sca} is a convex optimization problem and can be efficiently solved by standard interior-point methods through disciplined convex programming frameworks such as CVX.

\subsubsection{Passive STAR-RIS Coefficient Optimization}
\label{subsec:RIS_SCA}

In this block, with the active communication beamformers $\{\mathbf v_k\}_{k=1}^{K}$ and sensing beamformer $\mathbf v_0$ obtained in the previous optimization block, we then optimize the STAR-RIS coefficients. The resultant subproblem remains non-convex due to the coupled ES constraints and the quadratic dependence of the SINR expressions on the STAR-RIS coefficients.

\paragraph{Linear Representation of the Cascaded Channel}
Define
\(
\mathbf B_k
\triangleq
\operatorname{diag}(\mathbf h_k^*)\mathbf G
\in
\mathbb C^{N\times M_b}.
\) Using the identity
\(
\mathbf a^H\operatorname{diag}(\mathbf b)
=
\mathbf b^T\operatorname{diag}(\mathbf a^*),
\)
the cascaded STAR-RIS channel can be rewritten as
\begin{equation}
\label{eq:cascaded_linear}
\mathbf h_k^H\boldsymbol{\Phi}_i\mathbf G
=
\mathbf q_i^T\mathbf B_k.
\end{equation}

For each UE $k$ and stream index $\ell\in\{0,1,\ldots,K\}$, define
\(
s_{k,\ell}
\triangleq
\mathbf g_k^H\mathbf v_\ell,
\)
and
\(
\mathbf t_{k,\ell,i}
\triangleq
\mathbf B_k\mathbf v_\ell,
\quad
i\in\{t,r\}.
\)
Here, $\ell=0$ corresponds to the sensing stream. 
For UE $k$ located in region $i=i(k)$, the desired signal term is expressed as
\begin{equation}
z_k(\mathbf q_i)
=
s_{k,k}
+
\mathbf q_i^T\mathbf t_{k,k,i},
\end{equation}
while the interference-plus-noise term is given by
\begin{equation}
\psi_k(\mathbf q_i)
=
\sum_{\ell\neq k}
|s_{k,\ell}+\mathbf q_i^T\mathbf t_{k,\ell,i}|^2
+
|s_{k,0}+\mathbf q_i^T\mathbf t_{k,0,i}|^2
+
\sigma_k^2.
\end{equation}

\paragraph{SCA-Based Surrogate Construction}

To address the non-convexity w.r.t $\mathbf q_i$, SCA constructs a concave lower bound around the local point $\mathbf q_i^{(t-1)}$. Let
\[
\bar z_k=z_k(\mathbf q_i^{(t-1)}),\;
\bar\psi_k=\psi_k(\mathbf q_i^{(t-1)}),\;
s_k^{(t-1)}=\frac{|\bar z_k|^2}{\bar\psi_k},
\]
with
$
c_{1k}=\log_2(1+s_k^{(t-1)}),\;
{
c_{2k}
=
{
s_k^{(t-1)}
}{/[
(1+s_k^{(t-1)})\ln 2]
},
}$ and $
c_{3k}=|\bar z_k|^2.
$

Since $|z_k(\mathbf q_i)|^2$ is a convex quadratic function of $\mathbf q_i$, its first-order Taylor expansion yields
\begin{align}
    |z_k(\mathbf q_i)|^2
\ge
2\Re\!\left\{
\bar z_k^*
z_k(\mathbf q_i)
\right\}
-
|\bar z_k|^2.
\end{align}

On the other hand, we denote
\(
\widetilde z_k(\mathbf q_i)
\triangleq
2\Re\!\left\{
\bar z_k^*
z_k(\mathbf q_i)
\right\}
-
c_{3k}
\) and observe that for $\widetilde z_k(\mathbf q_i)>0$, the following inequality holds
\begin{equation}
\frac{c_{3k}}{|z_k(\mathbf q_i)|^2}
\le
\frac{c_{3k}}{\widetilde z_k(\mathbf q_i)}.
\end{equation}

Accordingly, a concave lower-bound approximation of the achievable rate is given by
\begin{equation}
\label{eq:ris_rate_lb}
\widehat R_k(\mathbf q_i)
\ge
c_{1k}
+
c_{2k}
\left[
2
-
\frac{c_{3k}}{\widetilde z_k(\mathbf q_i)}
-
\frac{\psi_k(\mathbf q_i)}{\bar\psi_k}
\right].
\end{equation}
Herein, since $\psi_k(\mathbf q_i)$ is convex quadratic and $\widetilde z_k(\mathbf q_i)$ is affine, the surrogate function $\widehat R_k(\mathbf q_i)$ is concave over the feasible region $\widetilde z_k(\mathbf q_i)>0$. Moreover, this also sheds light on the fact:
\[
\widehat R_k(\mathbf q_i^{(t-1)})
=
R_k(\mathbf q_i^{(t-1)}),
\]
which ensures local tightness of the surrogate approximation.

\paragraph{Convex STAR-RIS Subproblem}


\begingroup

Using the same ES penalty function and constraint introduced in
Section~IV-B, the ideal-hardware passive STAR-RIS
subproblem at iteration $i$ can be mathematically reformulated as
\begin{subequations}
\label{prob:sub2_sca_ideal}
\begin{align}
\max_{\mathbf q_t,\mathbf q_r,\tau}
\quad
&
\tau
+
\mu
\widehat f_{\mathrm{ES}}^{(i)}
(\mathbf q_t,\mathbf q_r)
\label{prob:sub2_sca_ideal_obj}
\\
\text{s.t.}\quad
&
|[\mathbf q_t]_n|^2
+
|[\mathbf q_r]_n|^2
\leq
1,
\quad
\forall n\in\mathcal N,
\label{prob:sub2_sca_ideal_es}
\\
&
\sum_{k=1}^{K}
\eta_k
\widehat R_k(\mathbf q_{i(k)})
\geq
\tau,
\label{prob:sub2_sca_ideal_wsr}
\\
&
\widehat R_k(\mathbf q_{i(k)})
\geq
r_{\mathrm{qos}},
\quad
\forall k\in\mathcal K,
\label{prob:sub2_sca_ideal_qos}
\\
&
\widetilde z_k(\mathbf q_{i(k)})
\geq
\epsilon,
\quad
\forall k\in\mathcal K,
\label{prob:sub2_sca_ideal_trust}
\\
&
\underline S^{(i)}
(\mathbf q_t,\mathbf q_r)
\geq
\delta_{\mathrm{ES}}.
\label{prob:sub2_sca_ideal_penalty}
\end{align}
\end{subequations}
where $\mu>0$ denotes the ES penalty coefficient and
$\delta_{\mathrm{ES}}>0$ guarantees that the reciprocal
penalty surrogate is well defined.

\endgroup

\begingroup

\begin{proposition}
\label{prop:ideal_penalty_convex}
For a fixed penalty coefficient $\mu>0$ and a given SCA
operating point
$\{\mathbf q_t^{(i-1)},\mathbf q_r^{(i-1)}\}$,
Problem~\eqref{prob:sub2_sca_ideal} is a convex optimization
problem. With the relaxed ES constraint
\eqref{prob:sub2_sca_ideal_es},
$f_{\mathrm{ES}}(\mathbf q_t,\mathbf q_r)\leq0$,
with equality if and only if the original element-wise
equality ES constraints are satisfied.
\end{proposition}

\begin{proof}
Under \eqref{prob:sub2_sca_ideal_es}, the inequality
\(
|[\mathbf q_t]_n|^2+|[\mathbf q_r]_n|^2\leq1,
\forall n\in\mathcal N,
\) implies
\(S(\mathbf q_t,\mathbf q_r)\leq N.\)
Hence,
\[
f_{\mathrm{ES}}(\mathbf q_t,\mathbf q_r)
=
{1}/{N}
-
{1}/{S(\mathbf q_t,\mathbf q_r)}
\leq0.
\]
Equality holds if and only if
$S(\mathbf q_t,\mathbf q_r)=N$.
Since each term contributing to
$S(\mathbf q_t,\mathbf q_r)$ is upper bounded by one,
this condition is equivalent to
\(
|[\mathbf q_t]_n|^2
+
|[\mathbf q_r]_n|^2
=
1,
\forall n\in\mathcal N.\)

We next establish the convexity of
Problem~\eqref{prob:sub2_sca_ideal}.
The function
$\underline S^{(i)}(\mathbf q_t,\mathbf q_r)$
is affine. Since $-1/x$ is concave for $x>0$,
$\widehat f_{\mathrm{ES}}^{(i)}
(\mathbf q_t,\mathbf q_r)$
is concave over the domain specified by
\eqref{prob:sub2_sca_ideal_penalty}.
Moreover,
$\widehat R_k(\mathbf q_{i(k)})$
is concave over
$\widetilde z_k(\mathbf q_{i(k)})>0$.
Therefore,
\(\tau
+
\mu
\widehat f_{\mathrm{ES}}^{(i)}
(\mathbf q_t,\mathbf q_r)\)
is a concave objective function.

Constraint \eqref{prob:sub2_sca_ideal_es}
is convex, while
\eqref{prob:sub2_sca_ideal_wsr}
and
\eqref{prob:sub2_sca_ideal_qos}
are superlevel sets of concave functions.
Furthermore,
\eqref{prob:sub2_sca_ideal_trust}
and
\eqref{prob:sub2_sca_ideal_penalty}
are affine constraints.
Hence,
Problem~\eqref{prob:sub2_sca_ideal}
maximizes a concave objective over a convex feasible set
and is therefore a convex optimization problem.
\end{proof}

\endgroup
\begingroup

For notational convenience, we define the corresponding  weight-sum rate and penalized objective  as
\begin{align}
    F_{\mathrm I}
(\mathbf v,\mathbf q_t,\mathbf q_r)
&\triangleq
\sum_{k\in\mathcal K}
\eta_k
R_k
(\mathbf v,\mathbf q_{i(k)}),
\\
\mathcal J_{\mathrm I,\mu}
(\mathbf v,\mathbf q_t,\mathbf q_r)
&\triangleq
F_{\mathrm I}
(\mathbf v,\mathbf q_t,\mathbf q_r)
+
\mu
f_{\mathrm{ES}}
(\mathbf q_t,\mathbf q_r).
\end{align}

\endgroup
\begin{algorithm}[ht]

\caption{{Proposed AO-based SCA Algorithm for Ideal ISAC Design}}
\label{alg:AO_SCA_ideal}

{
\begin{algorithmic}[1]

\State \textbf{Input:}
$\mathbf G$, $\{\mathbf h_k\}$, $\{\mathbf g_k\}$,
$\theta$,
$P_{\max}$,
$r_{\mathrm{qos}}$,
$\epsilon$,
$\mu>0$, and
$\varepsilon_{\mathrm{ES}}>0$.

\State \textbf{Initialize:}
Feasible points
$\{\mathbf v_k^{(0)}\}$,
$\mathbf v_0^{(0)}$,
and 
$\{\mathbf q_t^{(0)},\mathbf q_r^{(0)}\}$.
Set iteration index $i=0$.

\State Compute the initial penalized objective $\mathcal J_{\mathrm I,\mu}^{(0)}$.

\Repeat

\State For fixed
$\{\mathbf q_t^{(i)},\mathbf q_r^{(i)}\}$,
update
$\{\mathbf v_k^{(i+1)}\}$ and
$\mathbf v_0^{(i+1)}$
by solving Problem~\eqref{prob:active_sca}.

\State For fixed
$\{\mathbf v_k^{(i+1)}\}$ and
$\mathbf v_0^{(i+1)}$,
update
$\{\mathbf q_t^{(i+1)},\mathbf q_r^{(i+1)}\}$
by solving the penalized problem in
\eqref{prob:sub2_sca_ideal}.

\State Evaluate the ES equality residual
\[
e_{\mathrm{ES}}^{(i+1)}
=
\max_{n\in\mathcal N}
|
|[\mathbf q_t^{(i+1)}]_n|^2
+
|[\mathbf q_r^{(i+1)}]_n|^2
-
1
|.
\]

\State Evaluate the penalized objective $\mathcal J_{\mathrm I,\mu}^{(i+1)}$.

\State Compute the relative objective variation
\[
\Delta_{\mathrm I}^{(i+1)}
=
{
|
\mathcal J_{\mathrm I,\mu}^{(i+1)}
-
\mathcal J_{\mathrm I,\mu}^{(i)}
|
}/{
\max
\{
1,
|
\mathcal J_{\mathrm I,\mu}^{(i)}
|
\}
}.
\]

\State Set $i\leftarrow i+1$.

\Until{
$\Delta_{\mathrm I}^{(i)}
\leq
\epsilon$
and
$e_{\mathrm{ES}}^{(i)}
\leq
\varepsilon_{\mathrm{ES}}$
}

\State \textbf{Output:}
Optimized beamformers
$\{\mathbf v_k^\star\}$,
$\mathbf v_0^\star$,
and STAR-RIS coefficients
$\{\mathbf q_t^\star,\mathbf q_r^\star\}$.

\end{algorithmic}
}
\end{algorithm}

\subsubsection{Algorithm, Convergence and Complexity Analysis}
\label{subsec:convergence_complexity_ideal}

This subsection summarizes the iterative operation of the
ideal-hardware AO-based SCA algorithm and subsequently analyzes its
convergence behavior and computational complexity.
{
\paragraph{AO-Based Iterative Procedure}
\begingroup

The overall ideal-hardware AO-based SCA procedure is
summarized in Algorithm~\ref{alg:AO_SCA_ideal}. Initially, we generate the STAR-RIS coefficients $\boldsymbol{\Phi}_i^{(0)}$ within the constraints of \eqref{eq:es_const} and then derive the beamformers through jointly optimizing $(\{\mathbf v_k\},\,\mathbf v_0,\,\alpha)$ to maximize the minimum normalized feasibility margin $\alpha\ge 0$ such that  
\begin{align}
    \eqref{prob:wsr_main_power},
\frac{p(\theta)}{\Gamma_{\mathrm{tar}}}-1
\ge \alpha, \frac{
R_k(\{\mathbf v_l\},\boldsymbol{\Phi}_i^{(0)})
}{r_{\mathrm{qos}}}-1
\ge \alpha, \forall k\in\mathcal K.\nonumber
\end{align}
Having obtained these feasible points in hand, the algorithm alternately updates the active BS beamformers and passive STAR-RIS coefficients. In particular, for fixed
$\{\mathbf q_t^{(i)},\mathbf q_r^{(i)}\}$,
the communication beamformers
$\{\mathbf v_k\}$ and sensing beamformer $\mathbf v_0$
are updated by solving
Problem~\eqref{prob:active_sca}.
The updated beamformers are subsequently fixed, while
$\mathbf q_t$ and $\mathbf q_r$ are updated by solving the
penalized passive STAR-RIS subproblem
\eqref{prob:sub2_sca_ideal}.

In this way, the ES equality is promoted directly through the penalty
term, while the residual
$e_{\mathrm{ES}}^{(i)}$
is monitored after every AO iteration.
The algorithm terminates when both
$\Delta_{\mathrm I}^{(i)}\leq\epsilon$
and
$e_{\mathrm{ES}}^{(i)}
\leq\varepsilon_{\mathrm{ES}}$.
The final solution therefore satisfies the original ES
constraint within the prescribed numerical tolerance.

\endgroup

\begingroup

\paragraph{Convergence Analysis}

Using the penalized ideal-hardware objective
$\mathcal J_{\mathrm I,\mu}$ defined above,
consider first the active-beamforming update for fixed
$\{\mathbf q_t^{(i)},\mathbf q_r^{(i)}\}$.
Since the ES penalty remains constant during this update and
the active SCA surrogate is a locally tight lower bound of the
ideal-hardware WSR, the solution
$\mathbf v^{(i+1)}$ satisfies
\begin{equation}
\label{eq:ideal_conv_active}
\mathcal J_{\mathrm I,\mu}
(
\mathbf v^{(i+1)},
\mathbf q_t^{(i)},
\mathbf q_r^{(i)}
)
\geq
\mathcal J_{\mathrm I,\mu}
(
\mathbf v^{(i)},
\mathbf q_t^{(i)},
\mathbf q_r^{(i)}
).
\end{equation}

For fixed $\mathbf v^{(i+1)}$, define the passive surrogate
objective as
\[
\widehat{\mathcal J}_{\mathrm I,q}^{(i)}
(\mathbf q_t,\mathbf q_r)
\triangleq
\sum_{k\in\mathcal K}
\eta_k
\widehat R_k(\mathbf q_{i(k)})
+
\mu
\widehat f_{\mathrm{ES}}^{(i)}
(\mathbf q_t,\mathbf q_r).
\]
By construction,
$\widehat R_k$ and
$\widehat f_{\mathrm{ES}}^{(i)}$
are global lower bounds of their corresponding original
functions and are locally tight at the current SCA point.
Hence, one can obtain the inequality
\begin{equation}
\label{eq:ideal_conv_lower}
\widehat{\mathcal J}_{\mathrm I,q}^{(i)}
(\mathbf q_t,\mathbf q_r)
\leq
\mathcal J_{\mathrm I,\mu}
(
\mathbf v^{(i+1)},
\mathbf q_t,\mathbf q_r
),
\end{equation}
with equality at
$\{\mathbf q_t^{(i)},\mathbf q_r^{(i)}\}$.
Since
$\{\mathbf q_t^{(i+1)},\mathbf q_r^{(i+1)}\}$
is obtained by solving
Problem~\eqref{prob:sub2_sca_ideal},
we have
\begin{equation}
\label{eq:ideal_conv_surrogate}
\widehat{\mathcal J}_{\mathrm I,q}^{(i)}
(
\mathbf q_t^{(i+1)},
\mathbf q_r^{(i+1)}
)
\geq
\widehat{\mathcal J}_{\mathrm I,q}^{(i)}
(
\mathbf q_t^{(i)},
\mathbf q_r^{(i)}
).
\end{equation}
Therefore, one can further obtain the inequality
\begin{equation}
\label{eq:ideal_conv_passive}
\mathcal J_{\mathrm I,\mu}
(
\mathbf v^{(i+1)},
\mathbf q_t^{(i+1)},
\mathbf q_r^{(i+1)}
)
\geq
\mathcal J_{\mathrm I,\mu}
(
\mathbf v^{(i+1)},
\mathbf q_t^{(i)},
\mathbf q_r^{(i)}
).
\end{equation}

Combining
\eqref{eq:ideal_conv_active}
and
\eqref{eq:ideal_conv_passive}
gives
\begin{equation}
\label{eq:ideal_conv_total}
\mathcal J_{\mathrm I,\mu}^{(i+1)}
\geq
\mathcal J_{\mathrm I,\mu}^{(i)}.
\end{equation}
That is to say, the proposed penalty-based ideal-hardware AO-SCA algorithm generates a non-decreasing sequence ${\mathcal J_{\mathrm I,\mu}^{(i)}}$. Since the transmit power is bounded by $P_{\max}$, the ideal-hardware WSR is upper bounded. Moreover, $f_{\mathrm{ES}}(\mathbf q_t,\mathbf q_r)\leq0$ over the relaxed ES feasible set, ensuring that $\mathcal J_{\mathrm I,\mu}$ is upper bounded and ${\mathcal J_{\mathrm I,\mu}^{(i)}}$ converges to a finite value. Furthermore, $\mu$ is selected sufficiently large to ensure $e_{\mathrm{ES}}\leq\varepsilon_{\mathrm{ES}}$ at convergence.



\endgroup
}
{
\paragraph{Computational Complexity Analysis}
As in the HWI-aware design, the ideal-hardware algorithm solves the active and passive convex subproblems using interior-point methods \cite{8359418}. Following that,
the active and passive subproblems involve approximately $n_{1,\mathrm I}=2M_b(K+1)$ and $n_{2,\mathrm I}=4N+1$ real-valued variables, respectively, with corresponding constraint counts $m_{1,\mathrm I}$ and $m_{2,\mathrm I}$. Therefore, over $I_{\mathrm{AO}}$ AO iterations, the overall complexity is $\mathcal O\!\left[I_{\mathrm{AO}}\left(n_{1,\mathrm I}^{2}m_{1,\mathrm I}^{2.5}+m_{1,\mathrm I}^{3.5}+n_{2,\mathrm I}^{2}m_{2,\mathrm I}^{2.5}+m_{2,\mathrm I}^{3.5}\right)\right]$, 
which scales polynomially with the numbers of BS antennas, served UEs, and STAR-RIS elements for a given number of AO iterations.

}

\section{Simulation Results}
\label{sec4}

\subsection{Simulation Setup}

\begin{figure*}[!t]
    \centering
    \begin{minipage}[t]{0.48\textwidth}
        \centering
        \includegraphics[width=\linewidth]{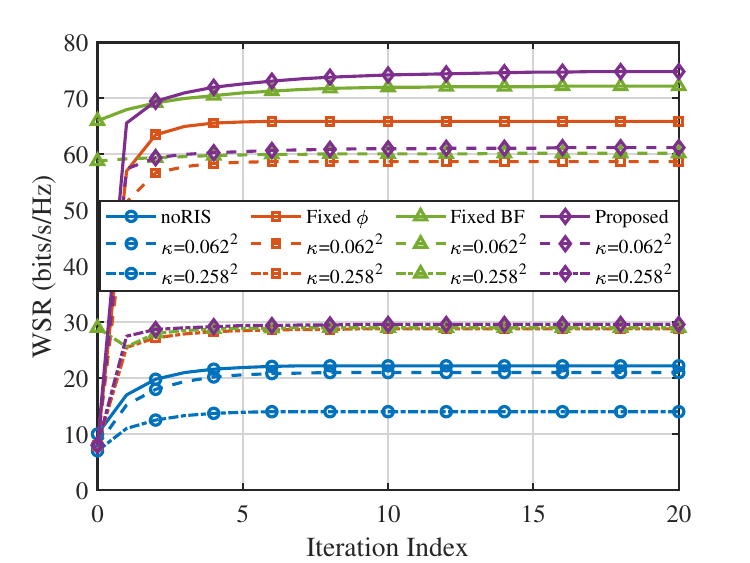}
        \caption{Convergence behavior of the proposed algorithms.}
        \label{fig:conv_p30g10}
    \end{minipage}\hfill%
    \begin{minipage}[t]{0.48\textwidth}
        \centering
        \includegraphics[width=\linewidth]{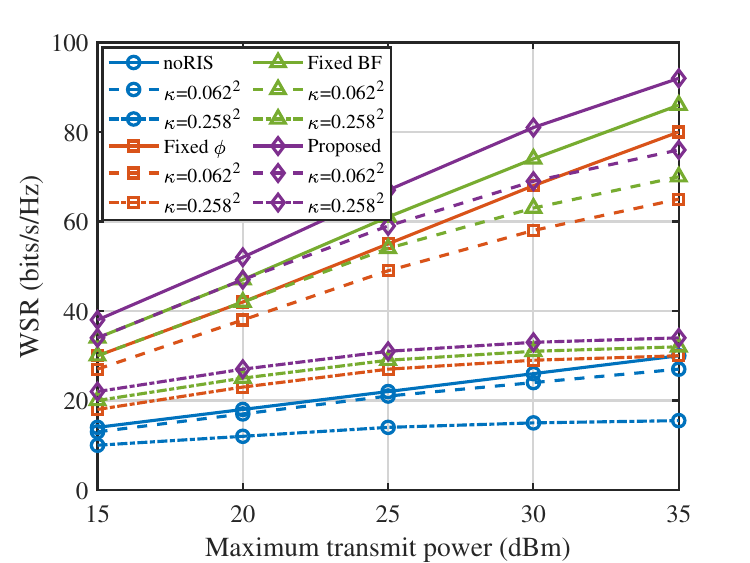}
        \caption{Average WSR versus the transmit power of the BS.}
        \label{fig:wsr_vs_pmax}
    \end{minipage}
\end{figure*}


Unless otherwise specified, the simulation parameters are set as $P_{\max}=25$\,dBm, $\Gamma_{\mathrm{tar}}=15$\,dBm, $r_{\mathrm{qos}}=1.5$\,bps/Hz, $f_c=10$\,GHz, $\lambda_c=0.03$\,m, $L=16$, $\sigma^2=-110$\,dBm, $M_b=16$, $M=1$, and $K=4$, with $\mathcal K_t=\{1,2\}$, $\mathcal K_r=\{3,4\}$, and $\eta_k=1$, $\forall k$.
The STAR-RIS is deployed in the vicinity of the BS to provide controllable full-space coverage for user clusters located on its transmission and reflection sides. The BS and STAR-RIS are located at $(0,0,0)$ m and $(0,50,0)$ m, respectively. The UEs are distributed over circular regions centered at the STAR-RIS with radii $d_{1t}=d_{1r}=2$ m and $d_{2t}=d_{2r}=4$ m. 
For an
($8\times5$)
STAR-RIS and having $N=40$ elements operating at
$10$\,GHz,
the corresponding Rayleigh distance is approximately
$5$\,m.
Since all UEs are located within
$4$\,m
from the STAR-RIS, they operate in the near-field region.
The BS--STAR channel paths are deterministically generated by linearly sweeping directional cosines, and the sensing directions are set to $18^\circ$, $72^\circ$, $108^\circ$, and $162^\circ$. The path-loss model is given by $\beta=C_0({d}/{D_0})^{-\alpha}$, where $C_0=-30$\,dB, $D_0=1$\,m, and $\alpha=2.2$.

{
To the best of our knowledge, the joint WSR optimization of an
ES-based STAR-RIS-assisted ISAC system considering hybrid
far-/near-field propagation, sensing constraints, and practical HWIs has not been studied yet in the existing literature, as early elaborated in Table~\ref{tab:contrasting_table}.
}
Thus, to demonstrate the effectiveness of the proposed framework (\textbf{proposed}), we compare it with the following baseline schemes:
\begin{itemize}
\item
\textbf{noRIS:}
Set
$\boldsymbol{\Phi}_t=\mathbf0$
and
$\boldsymbol{\Phi}_r=\mathbf0$,
and optimize
$\{\mathbf v_k\}$
and
$\mathbf v_0$
using only the direct channels
$\{\mathbf g_k\}$.

{
\item
\textbf{Fixed BF:}
The active beamformers are fixed, while the STAR-RIS transmission
and reflection coefficients $\mathbf q_t$ and $\mathbf q_r$ are
optimized.

\item
\textbf{Fixed $\phi$:}
The STAR-RIS coefficients $\mathbf q_t$ and $\mathbf q_r$ are fixed,
while the communication beamformers $\{\mathbf v_k\}$ and sensing
beamformer $\mathbf v_0$ are optimized.}
\end{itemize}
All schemes here are considered with the same system parameters and constraints, where all convex subproblems are conducted in MATLAB using the CVX programming framework and solved by the SDPT3 solver \cite{grant2014cvx,tutuncu2003solving}. 




\begin{figure*}[!t] 
\centering \begin{minipage}{0.333\linewidth} 
\centering \includegraphics[ width=0.9\linewidth, trim=15 0 15 0, clip ]{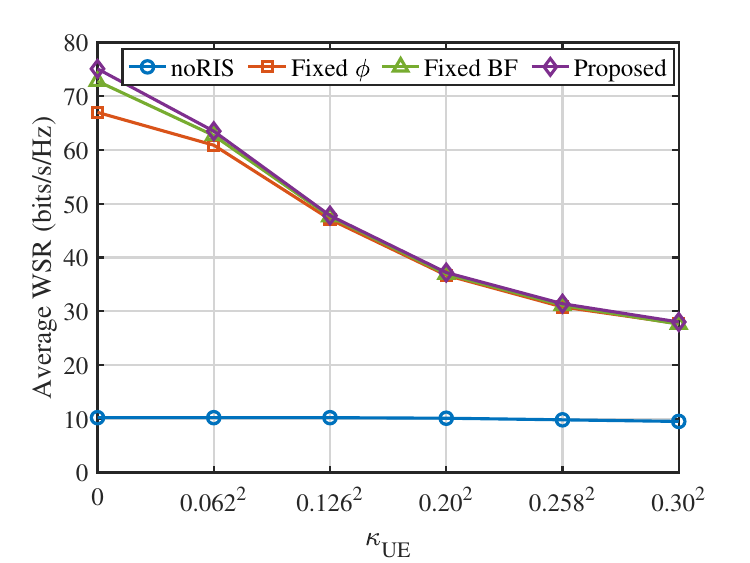}\\[-1mm] \small (a) \end{minipage}%
\begin{minipage}{0.333\linewidth}
\centering \includegraphics[ width=0.9\linewidth, trim=15 0 15 0, clip ]{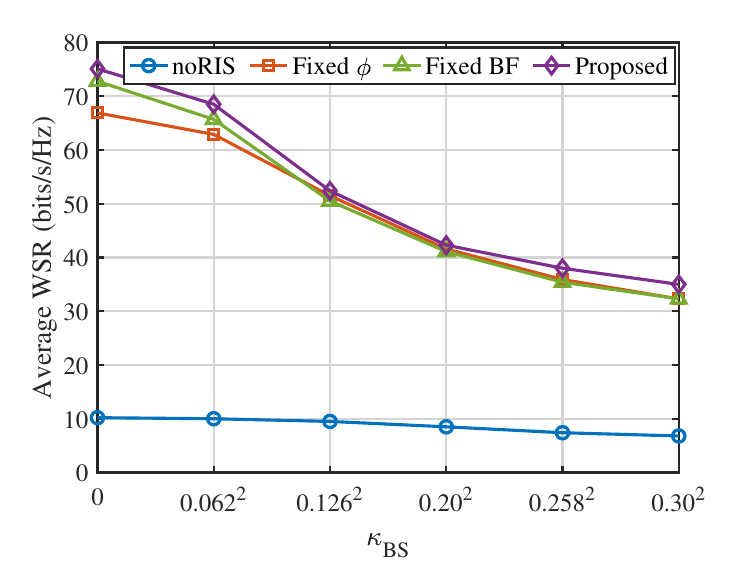}\\[-1mm] \small (b) 
\end{minipage}%
\begin{minipage}{0.333\linewidth} \centering \includegraphics[ width=0.9\linewidth, trim=15 0 15 0, clip ]{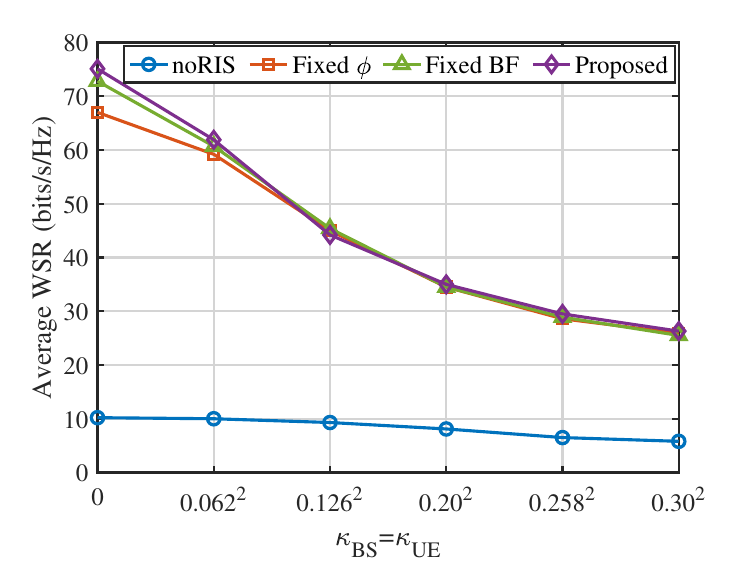}\\[-1mm] \small (c) 
\end{minipage} 

\caption{Average WSR versus the HWI coefficient $\kappa$: (a) UE-side HWI only; (b) BS-side HWI only; and (c) Joint BS/UE HWIs.} \label{fig:wsr_kappa}
\end{figure*}

\subsection{Performance Evaluation}

{
\subsubsection{Convergence Behavior}

Fig.~\ref{fig:conv_p30g10} illustrates the convergence behavior of
the schemes considered under different HWI levels. It can be observed
that all optimization schemes converge rapidly within a small number
of iterations. In particular, the proposed joint active/passive
optimization exhibits a monotonic increase in WSR and approaches a
stable solution after approximately $5$--$8$ iterations. The fixed
beamforming and fixed STAR-RIS schemes generally converge slightly
faster because only one optimization block is updated, whereas the
proposed method alternately optimizes both the BS beamformers and the STAR-RIS coefficients. Despite the larger
optimization space, the additional number of iterations required by the proposed scheme remains limited. The results also show that increasing the HWI level reduces convergent WSR, especially in the setting of severe impairment $\kappa^2=0.258^2$. Nevertheless, the convergence trajectories remain stable under all considered HWI conditions, indicating that the proposed AO-based SCA procedure is numerically robust to practical hardware distortion.
}

\begin{figure*}[!t]
    \centering
    \begin{minipage}[t]{0.48\textwidth}
        \centering
        \includegraphics[width=\linewidth]{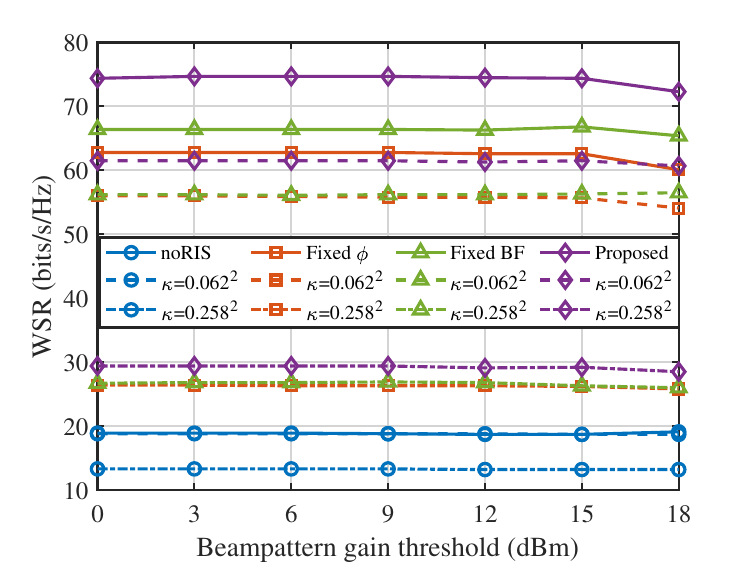}
        \caption{Average WSR versus beampattern gain threshold.}
        \label{fig:hwi_convergence_gamma}
    \end{minipage}\hfill%
    \begin{minipage}[t]{0.48\textwidth}
        \centering
        \includegraphics[width=\linewidth]{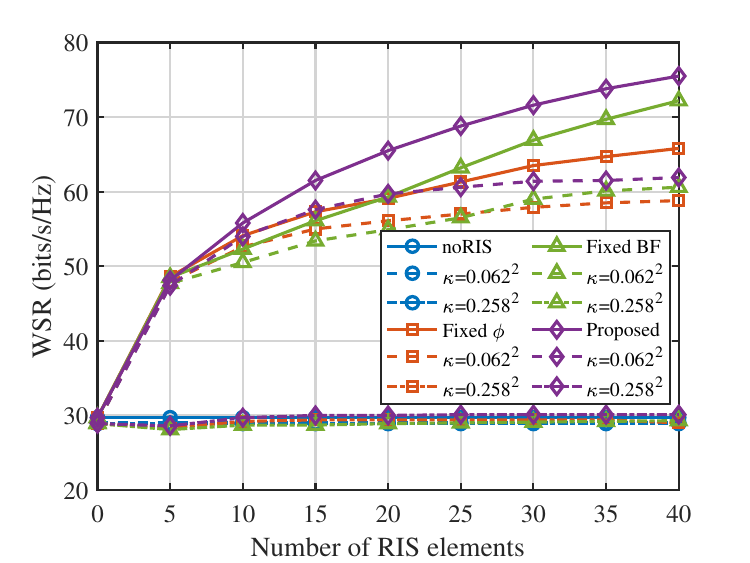}
        \caption{Average WSR versus STAR-RIS elements $N$ under HWIs.}
        \label{fig:hwi_convergence}
    \end{minipage}
\end{figure*}


\subsubsection{Impact of Maximum Transmit Power on WSR}
{
Fig.~\ref{fig:wsr_vs_pmax} shows the average WSR as a function of
the maximum transmit power $P_{\max}$ under different beamforming
strategies and HWI levels. In general, the WSR increases with
$P_{\max}$ since a larger power budget provides more flexibility for
joint communication and sensing transmission. Across all considered
settings, the proposed design achieves the
highest WSR, while the fixed-beamforming and fixed-STAR-RIS schemes
provide intermediate performance and the noRIS benchmark yields the
lowest WSR. This behavior highlights the advantage of jointly
optimizing the BS beamformers and the STAR-RIS transmission and
reflection coefficients. The effect of HWIs also becomes more evident as the
transmit power increases. With the mild impairment level
$\kappa^2=0.062^2$, the WSR continues to improve noticeably as
$P_{\max}$ increases. In contrast, under the severe impairment level
$\kappa^2=0.258^2$, the WSR gain becomes much smaller at higher
transmit powers. This is because the distortion power grows with the
transmit covariance, which limits the benefit of further increasing
$P_{\max}$. As a result, the system gradually becomes
distortion-limited in the high-power regime.
}
{
\subsubsection{Impact of Hardware Impairments on WSR}

Fig.~\ref{fig:wsr_kappa} illustrates the average WSR under different
HWI levels, considering UE-side impairments in
Fig.~\ref{fig:wsr_kappa}(a), BS-side impairments in
Fig.~\ref{fig:wsr_kappa}(b), and joint BS/UE impairments in
Fig.~\ref{fig:wsr_kappa}(c). In all cases, the WSR decreases as the
impairment coefficient increases. The proposed design achieves the highest WSR over most of the considered range,
followed closely by the fixed-beamforming and fixed-STAR-RIS schemes,
while the noRIS benchmark yields considerably lower performance. As the HWIs become more severe, the performance gap among the
STAR-RIS-assisted schemes gradually narrows because the
signal-dependent distortion increasingly limits the benefit of
beamforming optimization. This degradation is more pronounced when
the BS and UE impairments are jointly considered, as distortion is
introduced at both sides of the link. These results demonstrate the
importance of accounting for practical HWIs in STAR-RIS-assisted
ISAC design.
}
{
\subsubsection{Impact of Beampattern Gain Threshold on WSR}

Fig.~\ref{fig:hwi_convergence_gamma} shows the average WSR versus the
beampattern gain threshold $\Gamma_{\mathrm{tar}}$. The proposed
joint active/passive design achieves the highest WSR over the whole
range. At low and moderate sensing requirements, the WSR changes only
slightly, indicating that the sensing constraint does not yet strongly
restrict the communication-oriented beamforming solution. As
$\Gamma_{\mathrm{tar}}$ becomes more stringent, the achievable WSR
gradually decreases because more beamforming flexibility must be
devoted to satisfying the sensing requirement. HWIs further reduce
the achievable WSR, with the degradation becoming more evident under
severe impairment. These results clearly illustrate the trade-off
between sensing performance and communication throughput in the
considered STAR-RIS-assisted ISAC system.
}

\subsubsection{Impact of Hardware Specifications on WSR}
{
Figs.~\ref{fig:hwi_convergence} and \ref{fig:wsr_vs_Mb} illustrate
the effects of the STAR-RIS size $N$ and the number of BS antennas
$M_b$ on the achievable WSR, respectively. In general, increasing
either $N$ or $M_b$ improves the WSR by providing additional passive
or active beamforming degrees of freedom. A larger STAR-RIS offers
more flexibility in shaping the transmitted and reflected signals,
whereas increasing $M_b$ improves the spatial resolution and
beamforming capability at the BS. The proposed design achieves the highest WSR in both cases, while the
noRIS benchmark is almost insensitive to $N$ but still benefits from
a larger BS array. The influence of HWIs becomes more evident as the system dimension
increases. Under mild impairment levels, the WSR continues to improve
noticeably with both $N$ and $M_b$. In contrast, under severe HWIs,
the corresponding gains become much smaller because the
signal-dependent distortion increasingly offsets the additional
beamforming advantage provided by the larger surface or antenna
array. These results indicate that increasing the STAR-RIS size and
the number of BS antennas remains beneficial, but the achievable
scaling gain is ultimately constrained by the underlying hardware
quality.}

\subsubsection{Impact of HWIs on beampattern gain} 
\begin{figure*}[!t]
    \centering
    \begin{minipage}[t]{0.48\textwidth}
        \centering
        \includegraphics[width=\linewidth]{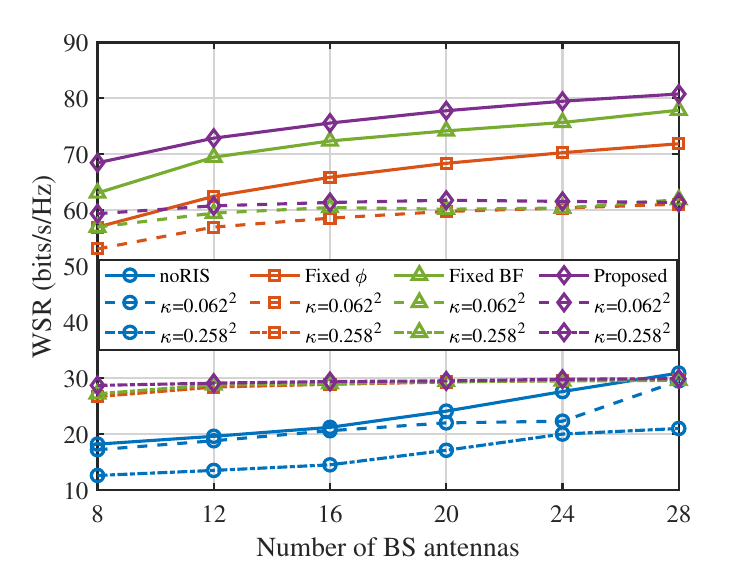}
        \caption{Average WSR versus the number of BS antennas $M_b$ under HWIs.}
        \label{fig:wsr_vs_Mb}
    \end{minipage}\hfill%
    \begin{minipage}[t]{0.48\textwidth}
        \centering
        \includegraphics[width=\linewidth]{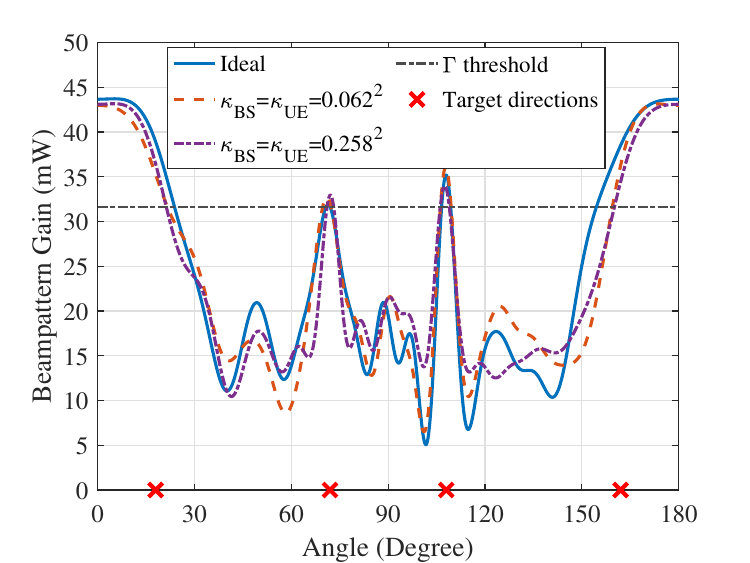}
        \caption{Beampattern gain versus angle.}
        \label{fig:beampatternGain}
    \end{minipage}
\end{figure*}

Fig.~\ref{fig:beampatternGain} shows the radar beampattern under different HWI levels, where the red crosses indicate the prescribed sensing directions. The beampattern gains at all target directions remain above the required threshold. Distinct peaks are formed around the intermediate target directions, while the targets near the angular boundaries lie within broader high-gain regions. This behavior arises because the sensing constraint imposes only a minimum gain requirement at each target direction, rather than forcing every target to coincide with an identical local maximum. The three curves remain close because the HWI-aware beamforming design readjusts the transmit covariance to preserve the sensing constraints. Increasing the HWI level mainly causes small variations in the peak and sidelobe levels, rather than a substantial change in the overall beampattern. Therefore, the proposed design maintains the required sensing performance even under severe HWIs.

\begin{figure*}[!t]
    \centering
    \begin{minipage}[t]{0.48\textwidth}
        \centering
        \includegraphics[width=\linewidth]{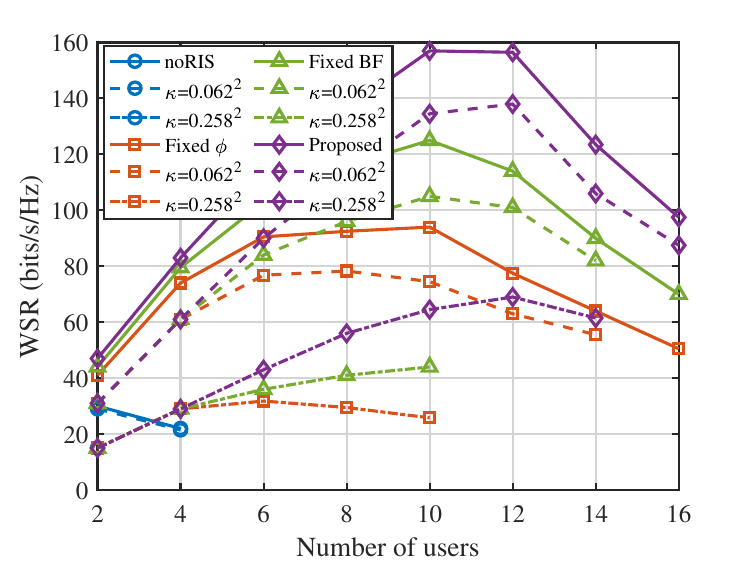}
        \caption{Average WSR versus the number of UEs.}
        \label{fig:wsr_vs_ues}
    \end{minipage}\hfill%
    \begin{minipage}[t]{0.48\textwidth}
        \centering
        \includegraphics[width=\linewidth]{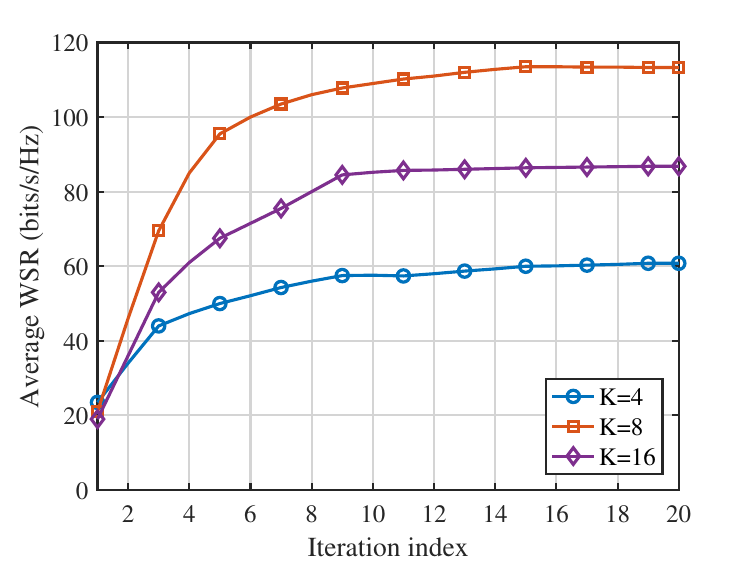}
        \caption{Convergence behavior for different numbers of UEs.}
        \label{fig:convergence_ues}
    \end{minipage}
\end{figure*}

{
\subsubsection{Impact of the Number of UEs on WSR and Convergence}

Fig.~\ref{fig:wsr_vs_ues} evaluates the average WSR as the number
of served UEs increases. The proposed joint design
generally achieves the highest WSR and maintains a clear advantage
over the fixed-beamforming, fixed-STAR-RIS, and noRIS baselines across
the considered user loads. As $K$ increases, the system must
simultaneously accommodate more communication streams, multi-user
interference, and QoS constraints under the same transmit-power and
sensing requirements. Consequently, some baseline configurations
become infeasible at high user loads, as indicated by the terminated
curves, whereas the proposed joint optimization remains feasible over
a broader range of $K$. This demonstrates the benefit of jointly
exploiting the active and passive beamforming degrees of freedom when
the system becomes more heavily loaded. The convergence results in
Fig.~\ref{fig:convergence_ues} further show that the proposed AO-SCA
algorithm retains stable convergence for $K=4$, $8$, and $16$ UEs.
Although increasing $K$ enlarges the optimization dimension and the
computational burden of each iteration, all considered user loads
reach a stable objective value within a moderate number of AO
iterations. These results support the scalability of the proposed
framework w.r.t increasing multi-user loading.
}

\section{Conclusions}\label{sec5}

The WSR optimization of a STAR-RIS-assisted multi-UE MIMO ISAC system was carried out under the ES protocol and practical HWIs. A hybrid-field channel model was considered, where the BS-STAR-RIS link follows far-field propagation and the STAR-RIS-UE links are modeled using near-field spherical wavefronts. The WSR maximization problem was formulated that jointly optimizes the active BS beamformers and passive STAR-RIS coefficients under realistic transmit-power, rate, sensing beampattern, and ES constraints. To address the resulting non-convexity, an AO-based SCA algorithm was developed by constructing tractable convex surrogate subproblems for the active and passive beamforming designs. Our numerical results showed that HWIs can cause WSR saturation in the high transmit-power regime, while stricter sensing requirements reduce the achievable communication throughput. The results also demonstrated that increasing the STAR-RIS aperture and the number of BS antennas improves the WSR, although the gain becomes limited under strong hardware distortion.

\balance
\bibliographystyle{IEEEtran}
\bibliography{IEEEabrv,reference}

\end{document}